\PassOptionsToPackage{unicode}{hyperref}
\PassOptionsToPackage{hyphens}{url}
\PassOptionsToPackage{dvipsnames,svgnames,x11names}{xcolor}
\documentclass[12pt]{article}

\usepackage{amsmath,amssymb}
\usepackage{iftex}
\usepackage{adjustbox}
\usepackage{bm}
\usepackage{algorithm}
\usepackage[noend]{algpseudocode}
\usepackage{multirow}      
\usepackage{dcolumn}       
\usepackage{makecell}      
\usepackage{rotating}      
\usepackage{tablefootnote}
\usepackage{subfigure}
\newcommand{\change}[1]{#1}

\ifPDFTeX
  \usepackage[T1]{fontenc}
  \usepackage[utf8]{inputenc}
  \usepackage{textcomp} 
\else 
  \usepackage{unicode-math}
  \defaultfontfeatures{Scale=MatchLowercase}
  \defaultfontfeatures[\rmfamily]{Ligatures=TeX,Scale=1}
\fi
\usepackage{lmodern}
\ifPDFTeX\else  
\fi
\IfFileExists{upquote.sty}{\usepackage{upquote}}{}
\IfFileExists{microtype.sty}{
  \usepackage[]{microtype}
  \UseMicrotypeSet[protrusion]{basicmath} 
}{}
\makeatletter
\@ifundefined{KOMAClassName}{
  \IfFileExists{parskip.sty}{%
    \usepackage{parskip}
  }{
    \setlength{\parindent}{0pt}
    \setlength{\parskip}{6pt plus 2pt minus 1pt}}
}{
  \KOMAoptions{parskip=half}}
\makeatother
\usepackage{xcolor}
\makeatletter
\ifx\paragraph\undefined\else
  \let\oldparagraph\paragraph
  \renewcommand{\paragraph}{
    \@ifstar
      \xxxParagraphStar
      \xxxParagraphNoStar
  }
  \newcommand{\xxxParagraphStar}[1]{\oldparagraph*{#1}\mbox{}}
  \newcommand{\xxxParagraphNoStar}[1]{\oldparagraph{#1}\mbox{}}
\fi
\ifx\subparagraph\undefined\else
  \let\oldsubparagraph\subparagraph
  \renewcommand{\subparagraph}{
    \@ifstar
      \xxxSubParagraphStar
      \xxxSubParagraphNoStar
  }
  \newcommand{\xxxSubParagraphStar}[1]{\oldsubparagraph*{#1}\mbox{}}
  \newcommand{\xxxSubParagraphNoStar}[1]{\oldsubparagraph{#1}\mbox{}}
\fi
\makeatother

\usepackage{longtable,booktabs,array}
\usepackage{calc} 
\usepackage{etoolbox}
\makeatletter
\patchcmd\longtable{\par}{\if@noskipsec\mbox{}\fi\par}{}{}
\makeatother
\IfFileExists{footnotehyper.sty}{\usepackage{footnotehyper}}{\usepackage{footnote}}
\makesavenoteenv{longtable}
\usepackage{graphicx}
\makeatletter
\def\maxwidth{\ifdim\Gin@nat@width>\linewidth\linewidth\else\Gin@nat@width\fi}
\def\maxheight{\ifdim\Gin@nat@height>\textheight\textheight\else\Gin@nat@height\fi}
\makeatother
\setkeys{Gin}{width=\maxwidth,height=\maxheight,keepaspectratio}
\makeatletter
\def\fps@figure{htbp}
\makeatother

\makeatletter
\@ifpackageloaded{caption}{}{\usepackage{caption}}
\AtBeginDocument{%
\ifdefined\contentsname
  \renewcommand*\contentsname{Table of contents}
\else
  \newcommand\contentsname{Table of contents}
\fi
\ifdefined\listfigurename
  \renewcommand*\listfigurename{List of Figures}
\else
  \newcommand\listfigurename{List of Figures}
\fi
\ifdefined\listtablename
  \renewcommand*\listtablename{List of Tables}
\else
  \newcommand\listtablename{List of Tables}
\fi
\ifdefined\figurename
  \renewcommand*\figurename{Figure}
\else
  \newcommand\figurename{Figure}
\fi
\ifdefined\tablename
  \renewcommand*\tablename{Table}
\else
  \newcommand\tablename{Table}
\fi
}
\@ifpackageloaded{float}{}{\usepackage{float}}
\floatstyle{ruled}
\@ifundefined{c@chapter}{\newfloat{codelisting}{h}{lop}}{\newfloat{codelisting}{h}{lop}[chapter]}
\floatname{codelisting}{Listing}

\makeatother
\makeatletter
\@ifpackageloaded{caption}{}{\usepackage{caption}}
\@ifpackageloaded{subcaption}{}{\usepackage{subcaption}}
\makeatother

\ifLuaTeX
  \usepackage{selnolig}  
\fi
\usepackage[]{natbib}
\usepackage{bookmark}

\IfFileExists{xurl.sty}{\usepackage{xurl}}{} 
\hypersetup{
  pdftitle={Title},
  pdfauthor={Author 1; Author 2},
  pdfkeywords={3 to 6 keywords, that do not appear in the title},
  colorlinks=true,
  linkcolor={blue},
  filecolor={Maroon},
  citecolor={Blue},
  urlcolor={Blue},
  pdfcreator={LaTeX via pandoc}}

\counterwithout{lemma}{section}
\newtheorem{assumption}{Assumption}[section]
\counterwithout{assumption}{section}

\counterwithout{condition}{section}
\newtheorem{theorem}{Theorem}[section]
\counterwithout{theorem}{section}

\counterwithout{remark}{section}

\counterwithout{definition}{section}

\counterwithout{corollary}{section}
\newtheorem{proposition}{Proposition}[section]
\counterwithout{proposition}{section}
\numberwithin{equation}{section}
\definecolor{LightPink}{RGB}{255,228,225}
\DeclareMathOperator*{\argmin}{arg\,min}

\newcommand{\anon}{1}

\begin{document}

\def\spacingset#1{\renewcommand{\baselinestretch}%
{#1}\small\normalsize} \spacingset{1}


\if1\anon
{
  \title{\bf Pattern-Calibrated Multimodal Prediction under Blockwise Missingness}
  \author{Junhan Yu$^{1,*}$, Kejian Zhang$^{1,*}$, Doudou Zhou$^{1,*,\dagger}$, Guojun Zhu$^{2,*}$\\
    \small $^1$ Department of Statistics \& Data Science, National University of Singapore \\
    \small $^2$ School of Mathematical Sciences, University of Chinese Academy of Sciences \\
    \small $^*$ Alphabetical order\\
    \small $^\dagger$ Corresponding author: ddzhou@nus.edu.sg\\
    }
    \date{}
  \maketitle
} \fi

\if0\anon
{
  \bigskip
  \bigskip
  \bigskip
  \begin{center}
    {\LARGE\bf Pattern-Calibrated Multimodal Prediction under Blockwise Missingness}
\end{center}
  \medskip
} \fi

\bigskip
\begin{abstract}
Blockwise missingness in multimodal data creates observed-modality patterns with different prediction targets. A pooled rule need not be optimal for any fixed pattern, whereas pattern-specific fitting can be inefficient when some patterns are rare. Borrowing is especially difficult when two patterns share no raw modality and hence no directly comparable covariate component. We propose Multimodal Overlap-aware Shared-specific Alignment and Inter-pattern Calibration, a framework that learns aligned shared and modality-specific representations from partially observed multimodal data, then estimates a separate predictor for each pattern. For a focal pattern, observations from all patterns contribute only through representation components available under the focal pattern; focal-pattern observations then estimate the remaining correction. Under a linear latent-representation model, non-asymptotic bounds show how representation coverage, correction size, and representation error determine whether borrowing improves on local fitting. They also show that, when co-observed data align the modalities upstream, observations from raw-disjoint patterns can increase the effective sample size for estimating the shared component. Simulations, a naturally incomplete intensive-care mortality application, and controlled-masking studies of emotion recognition and glaucoma classification show that the method is most useful when focal-pattern samples are limited and pattern-specific discrepancies are appreciable.
\end{abstract}

\noindent%
\vfill

\newpage
\spacingset{1.8} 

\section{Introduction}\label{sec intro}

\begin{figure}[ht]
    \centering
    \includegraphics[width=0.85\textwidth]{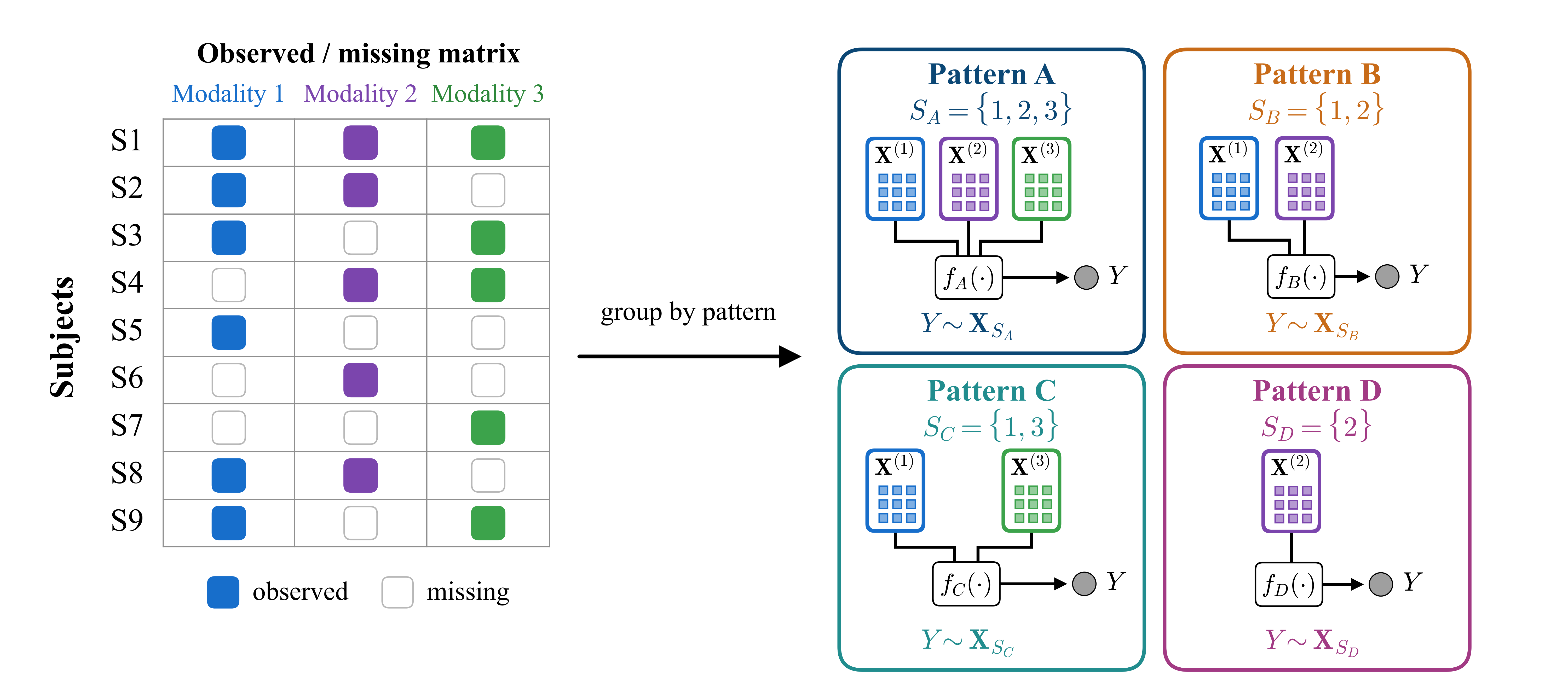}
    \caption{Blockwise missingness induces observed-modality patterns with different information sets. Each pattern defines the covariates available to its prediction rule.}
    \label{fig:missing-patterns}
\end{figure}

Prediction studies increasingly combine heterogeneous modalities, including structured records, text, images, and molecular assays \citep{baltrusaitis2019multimodal,kline2022multimodal,argelaguet2018multiomics}. Because these modalities arise from different acquisition systems and workflows, only a subset may be observed for each subject; in electronic health records (EHRs), for example, structured codes, clinical notes, and chest images may be available for different subsets of patients \citep{krones2025review}. This blockwise missingness induces observed-modality patterns, each defining the information available for prediction and therefore potentially inducing a different optimal prediction rule \citep{mercaldo2020missing,lemorvan2020linear}. A common pooled rule across patterns may therefore fail to recover the optimum for a given pattern, whereas fitting each pattern separately can be inefficient when pattern-specific samples are small. Figure~\ref{fig:missing-patterns} illustrates these pattern-specific prediction tasks.

Prior work has addressed pattern-dependent prediction through local fitting, mask-dependent modelling, and parameter sharing. Pattern submodels fit a separate prediction rule for each missingness pattern using only observations from that pattern \citep{mercaldo2020missing}. Mask-dependent Bayes prediction has been represented through interactions between observed values and missingness indicators, mask-aware neural architectures, and learned representations of the missingness pattern \citep{lemorvan2020linear,lemorvan2020neumiss,ma2025penn}. Sharing pattern submodels (SPSM) couple a common coefficient vector with sparse pattern-specific deviations, allowing information sharing without forcing every pattern to use the same rule \citep{stempfle2023sharing}. However, all three strategies express the pattern-specific predictors in a common covariate system.

A complementary line of work extracts predictive information from incomplete observations. Multimodal methods handle missing modalities by reconstructing them or by using zero-filling, missingness masks, reconstruction losses, and specialized fusion architectures \citep{tran2017missing,chartsias2018multimodal,wu2018multimodal,shi2019variational,yuan2021transformer,zuo2023exploiting,wu2026deep}. Related shared--specific methods separate representations common across modalities from those specific to each modality \citep{hazarika2020misa,lee2021private,wang2023shaspec}. Statistical methods borrow through prespecified structures: HAM uses observations containing the covariate subsets required by particular ANOVA components \citep{sell2024ham}, DISCOM pools moments across available modality pairs \citep{yu2020discom}, and Modular Regression uses designated auxiliary covariates or tasks \citep{jin2023modular}. Thus, these approaches provide either multimodal fusion and representation learning or borrowing through predefined covariate or task relationships. What remains unresolved is how to construct comparable representations for heterogeneous observed-modality patterns and borrow across them while retaining a distinct prediction rule for each pattern.

Extending pattern-aware borrowing to multimodal data therefore requires solving two linked problems. First, two patterns may share no raw covariates. In EHR data, for example, a code-only pattern and a note-only pattern provide no common covariate through which information can be borrowed. Borrowing between them therefore requires representations that make information extracted from different modalities comparable. Such representations can be learned from observations in which the relevant modalities are jointly observed, even when the two patterns themselves have no raw modality in common. Second, borrowing must preserve the prediction rule of every pattern rather than produce one global multimodal predictor. For a focal pattern, observations from another pattern can contribute only through representation components available under both patterns. Observations from the focal pattern must then be used to estimate the gap between the resulting preliminary predictor and the focal pattern's prediction rule.

To address these two problems, we propose \textbf{M}ultimodal \textbf{O}verlap-aware \textbf{S}hared-specific \textbf{A}lignment and \textbf{I}nter-pattern \textbf{C}alibration (\textbf{MOSAIC}). MOSAIC first learns aligned shared and modality-specific representations from partially observed multimodal observations without using response labels. For each pattern, it then forms a preliminary predictor using observations across all patterns, while retaining from each pattern only the representation components also available under the focal pattern. MOSAIC uses observations from the focal pattern to calibrate the preliminary predictor to that pattern's prediction rule. The preliminary predictor can therefore be estimated using more observations than a local fit, while the final predictor remains pattern-specific.

Our contributions are three-fold. First, we formulate blockwise multimodal prediction as a collection of pattern-specific risk minimization problems over potentially disjoint raw covariate spaces, thereby distinguishing each pattern's target from the average risk optimized by a pooled predictor. Second, we develop a unified estimation procedure that operates on modality-specific inputs and uses observations across patterns to estimate a distinct predictor for each pattern. Third, our non-asymptotic analysis characterizes how representation coverage, the gap between the preliminary predictor and the focal-pattern rule, and representation estimation error determine the benefit of borrowing. It identifies regimes in which MOSAIC improves on local fitting and shows that observations from patterns with no raw modality in common can increase the effective sample size for estimating the shared component of a pattern-specific predictor. Simulations and multimodal applications examine the roles of representation learning, cross-pattern borrowing, and pattern-specific calibration.

The rest of the paper is organized as follows. Section~\ref{sec methodology} presents the proposed method. Section~\ref{sec theory} establishes theoretical guarantees. Section~\ref{sec:simulation} reports simulation results; Section~\ref{sec real data} presents real-data analyses; and Section~\ref{sec conclusion} concludes.

\section{Methodology}\label{sec methodology}

\subsection{Problem Setup and Method Overview}\label{sec:statistical_formulation}

For subject $i$, let $y_i$ be the response and let $\mathbf x_i^{(l)}$ denote the raw input from modality $l\in[L]$, where $[L]=\{1,\ldots,L\}$. For any modality set $A\subseteq[L]$, write $\mathcal X_i(A)=\{\mathbf x_i^{(l)}:l\in A\}$, and let $\mathcal X(A)$ denote the corresponding generic multimodal observation. The complete multimodal observation $\mathcal X_i([L])$ may contain heterogeneous objects such as structured codes, text, or images. Let $m_i^{(l)}\in\{0,1\}$ indicate whether modality $l$ is observed, and define $\mathcal A_i=\{l\in[L]:m_i^{(l)}=1\}$, with $\mathcal A$ denoting the corresponding generic random set. The realized set $\mathcal A_i$ is subject $i$'s observed-modality pattern; we denote $K$ distinct nonempty values of $\mathcal{A}$ by $A_1,\ldots,A_K$.

Because the available covariates differ by pattern, the prediction target must first be defined for each pattern. Let $\mathfrak X_{A_k}$ denote the raw covariate space for observations with modalities $A_k$. For $A_k$, define the conditional risk and its raw-space minimizer as
\begin{equation}\label{eq:raw_pattern_target}
\begin{aligned}
    R_k^{\rm raw}(g)
    &:=\mathbb E\left[\{Y-g(\mathcal X(A_k))\}^2\mid \mathcal A=A_k\right],
    &g^{\rm raw}_{(k)}
    &:=\argmin_{g\in\mathcal G^{\rm raw}_k}R_k^{\rm raw}(g),
\end{aligned}
\end{equation}
where $\mathcal G^{\rm raw}_k$ is a class of prediction rules using the raw modalities in pattern $A_k$. This is the population target of pattern submodels \citep{mercaldo2020missing}; when $\mathcal G_k^{\rm raw}$ is unrestricted, it is the Bayes predictor conditional on the observed values and missingness pattern \citep{lemorvan2020linear}.

To contrast this target with pooling, let $\widetilde{\mathcal G}^{\rm raw}$ be a class of predictors $\widetilde g(x,A)$, defined for $A\in\{A_1,\ldots,A_K\}$ and $x\in\mathfrak X_A$. For every $\widetilde g\in\widetilde{\mathcal G}^{\rm raw}$ and $k\in[K]$, let $\widetilde g_k(x):=\widetilde g(x,A_k)$ and suppose $\widetilde g_k\in\mathcal G_k^{\rm raw}$. The pooled population rule is
\[
    \begin{aligned}
        \widetilde g^{\rm raw}_{\rm pool}&:=\argmin_{\widetilde g\in\widetilde{\mathcal G}^{\rm raw}}\mathbb E\left[\{Y-\widetilde g(\mathcal X(\mathcal A),\mathcal A)\}^2\right]\\
        &=\argmin_{\widetilde g\in\widetilde{\mathcal G}^{\rm raw}}\sum_{k=1}^K \pi_k\mathbb E\left[\{Y-\widetilde g(\mathcal X(A_k),A_k)\}^2\mid \mathcal A=A_k\right],
    \end{aligned}
\]
where $\pi_k=\mathrm{Pr}(\mathcal A=A_k)$.

\begin{proposition}[Pattern-specific excess risk of pooled rules]
\label{prop:pooling-target-failure}
Suppose that, for each $k\in[K]$, $\mathbb E(Y^2\mid \mathcal A=A_k)<\infty$ and $\mathcal G_k^{\rm raw}$ is closed and convex in $L_2(P_k)$, where $P_k$ is the conditional distribution of
$\mathcal X(A_k)$ given $\mathcal A=A_k$. Then, for the pooled population rule $\widetilde g^{\rm raw}_{\rm pool}$,
\[
    R_k^{\rm raw}(\widetilde g^{\rm raw}_{{\rm pool},k})-R_k^{\rm raw}(g^{\rm raw}_{(k)})\ge\left\|\widetilde g^{\rm raw}_{{\rm pool},k}-g^{\rm raw}_{(k)}\right\|_{L_2(P_k)}^2, \quad \forall k\in[K].
\]
\end{proposition}

Proposition~\ref{prop:pooling-target-failure} identifies the approximation cost of pooling: whenever the rule induced by the pooled model differs from $g^{\rm raw}_{(k)}$, pattern $A_k$ incurs positive excess risk. If the pooled class can realize all $K$ pattern-specific optima simultaneously, this cost vanishes. Let $n_k$ denote the number of labelled observations with pattern $A_k$ in a sample of size $n=\sum_{k=1}^K n_k$. A pattern-saturated class can attain all $K$ optima, but its empirical risk minimization separates into pattern-specific fits, each based only on $n_k$ observations. Imposing cross-pattern structure can reduce estimation error when some $n_k$ are small, but introduces approximation bias if the structure excludes any pattern-specific optimum. Useful borrowing must therefore link the pattern-specific rules without forcing them to coincide.

In multimodal data, constructing such a link is difficult because comparable predictor components may not exist in the raw covariate space. If $A_j\cap A_k=\varnothing$, observations from the two patterns contain different raw variables, so their raw-space predictors have no covariate-indexed coefficient or function component that can be matched and shared directly. Shared latent representations provide a way around this limitation by mapping different modalities into comparable representations \citep{lee2021private,wang2023shaspec}. These works, however, do not study how such representations can support statistical borrowing across pattern-specific prediction rules.

\change{To enable such borrowing, MOSAIC constructs a common latent coordinate system to represent the complete multimodal information. Formally, the complete multimodal information is represented by a shared representation $\mathbf z_i^{(0)}\in\mathbb R^{d^{(0)}}$ and modality-specific representations $\mathbf z_i^{(l)}\in\mathbb R^{d^{(l)}}$ for $l\in[L]$.} The raw modalities are generated from these representations as
\begin{equation}\label{eq:gen_model_x}
    \mathbf{x}_i^{(l)}
    = f_x^{(l)}(\mathbf{z}_i^{(0)}, \mathbf{z}_i^{(l)}, \bm{\varepsilon}_{x,i}^{(l)}),\quad l\in[L],
    \qquad \mathbf z_i=(\mathbf z_i^{(0)\top},\mathbf z_i^{(1)\top},\ldots,\mathbf z_i^{(L)\top})^\top\in\mathbb{R}^{d},
\end{equation}
where the $f_x^{(l)}$ are unknown functions, $\bm{\varepsilon}_{x,i}^{(l)}$ is the noise, and $d=d^{(0)}+\sum_{l=1}^L d^{(l)}$. Although $\mathbf z_i$ in \eqref{eq:gen_model_x} contains representation blocks associated with all modalities, an observation with pattern $A_k$ provides only the shared block $\mathbf z_i^{(0)}$ and the modality-specific blocks $\{\mathbf z_i^{(l)}:l\in A_k\}$. We therefore index the representations available under $A_k$ by $S_k=\{0\}\cup A_k$, where $0$ denotes the shared representation, and write their total dimension as $d_k=\sum_{l\in S_k}d^{(l)}$.

This alignment changes what can be borrowed. For a focal pattern $A_k$, every observation supplies an estimate of the same aligned shared block $\mathbf z^{(0)}$, regardless of which raw modalities produced it. An observation with pattern $A_j$ additionally supplies a modality-specific block $l$ required under $A_k$ only when $l\in A_j\cap A_k$. Consequently, even when $A_j\cap A_k=\varnothing$, observations with pattern $A_j$ still enter the preliminary prediction step through $\mathbf z^{(0)}$, which acts as a channel for information sharing despite the absence of common raw modalities. \change{This distinction is formalized in Proposition~\ref{prop:spsm-mosaic-borrowing}. Unlike prior work that shares coefficients through common raw modalities, MOSAIC can borrow through the shared representation even when two patterns have disjoint raw modalities.}

\begin{figure}[ht]
\centering
\includegraphics[width=\textwidth]{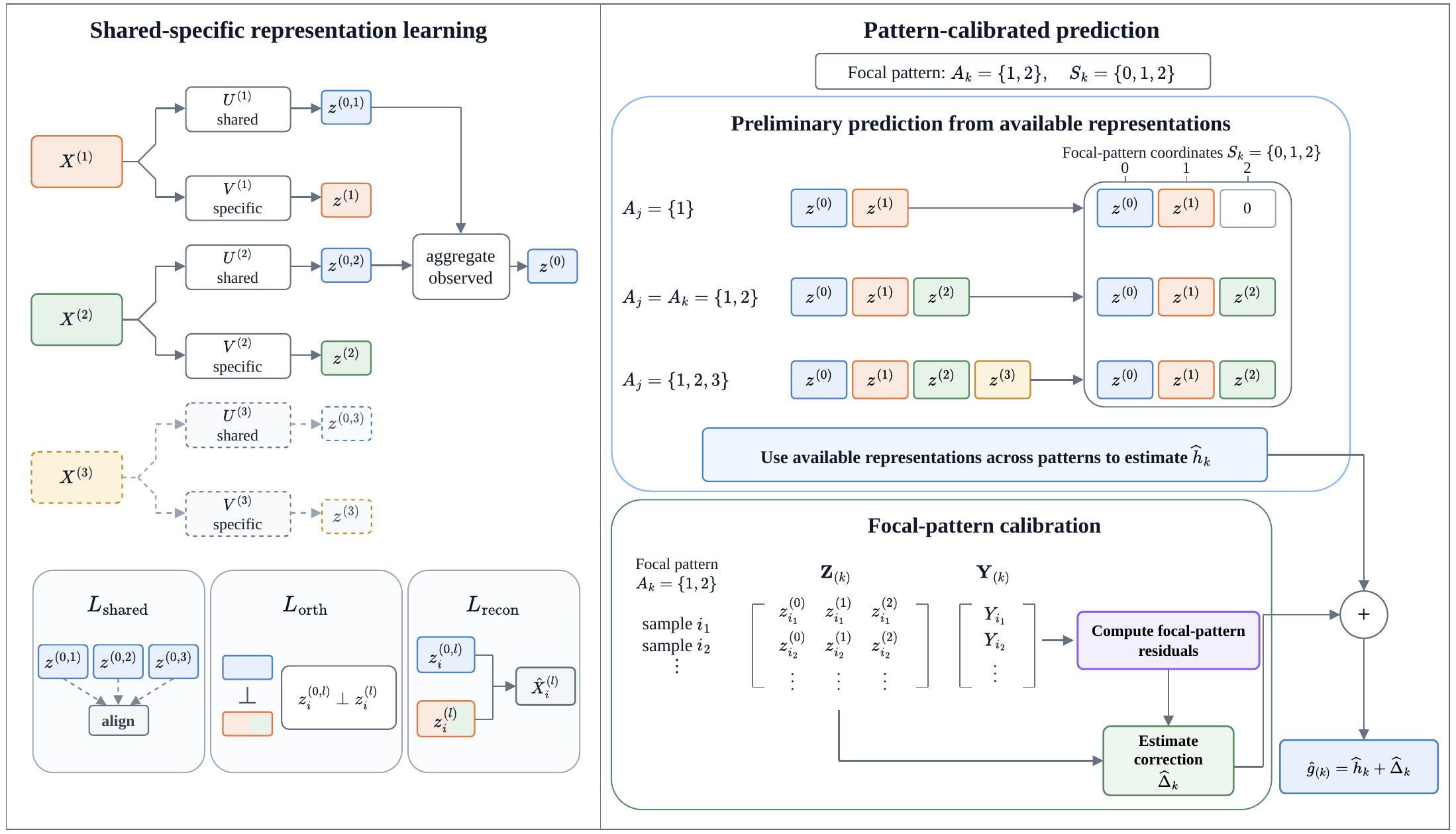}
\caption{Overview of MOSAIC in a three-modality example with focal pattern $A_k=\{1,2\}$ and representation index set $S_k=\{0,1,2\}$. MOSAIC learns aligned shared and modality-specific representations, constructs a preliminary predictor by retaining from each pattern only the representation components available under the focal pattern, and estimates the remaining correction from focal-pattern observations.}
\label{fig:framework-mosaic}
\end{figure}

The aligned representations provide fixed-dimensional inputs for standard prediction algorithms as well as the coordinates needed for cross-pattern borrowing. For pattern $A_k$, let $\mathbf z_{(k)}\in\mathbb R^{d_k}$ denote the concatenation of the representations indexed by $S_k$, and let $\mathcal G_k$ be a class of prediction functions from $\mathbb R^{d_k}$ to $\mathbb R$. We define the representation-space target as
\begin{equation}
    g_{(k)}:=\argmin_{g\in\mathcal G_k}\mathbb E\left[\{Y-g(\mathbf z_{(k)})\}^2\mid \mathcal A=A_k\right].
    \label{eq:pattern_specific_target}
\end{equation}
MOSAIC estimates a separate $g_{(k)}$ for each pattern. For a focal pattern $A_k$, it first uses observations from all patterns to estimate a preliminary predictor $h_k$ from the representation blocks that each pattern can supply. It then uses observations with pattern $A_k$ to estimate the remaining discrepancy. At the population level,
\[
    g_{(k)}=h_k+\Delta_k,
    \qquad \Delta_k:=g_{(k)}-h_k.
\]
Borrowing is beneficial when observations across patterns provide sufficient coverage to estimate $h_k$ accurately and $\Delta_k$ is easier to estimate from the $n_k$ focal-pattern observations than the full rule $g_{(k)}$. Figure~\ref{fig:framework-mosaic} summarizes these three stages of MOSAIC; the next three subsections define the representation learner, preliminary predictor, and focal-pattern calibration.

\subsection{Learning Representations for Cross-Pattern Borrowing}\label{sec:representation_learning}

We now describe how MOSAIC learns the aligned shared and modality-specific representations used for cross-pattern borrowing.

The representation-learning sample consists of $N$ multimodal observations indexed by $i=n+1,\ldots,n+N$, whose response labels are not used and may be unavailable. This separates it from the labelled prediction sample indexed by $i=1,\ldots,n$. After modality-specific preprocessing, each observed modality is represented as a vector in $\mathbb R^{d_{\mathrm{raw}}^{(l)}}$. For each modality $l\in[L]$, we learn a shared representation encoder $U^{(l)}:\mathbb R^{d_{\mathrm{raw}}^{(l)}}\to\mathbb R^{d^{(0)}}$ and a modality-specific representation encoder $V^{(l)}:\mathbb R^{d_{\mathrm{raw}}^{(l)}}\to\mathbb R^{d^{(l)}}$.

The shared encoders are aligned using observations in which two modalities are jointly observed. Since the shared representation participates in the generation of every modality, their shared encoder outputs for the same subject should be close. For pairs of distinct modalities with positive co-observation counts, we use:
\begin{equation}
    \mathcal{L}_{\mathrm{shared}} = \frac{1}{|\mathcal P_N|}\sum_{(l,l')\in\mathcal P_N} \frac{1}{\sum_{i=n+1}^{n+N} m_i^{(l)}m_i^{(l')}}\sum_{i=n+1}^{n+N} m_i^{(l)}m_i^{(l')} \left[ \ell_{\mathrm{align}}\left( U^{(l)}(\mathbf{x}_i^{(l)}), U^{(l')}(\mathbf{x}_i^{(l')}) \right) \right],
    \label{eq:rep_shared_loss}
\end{equation}
where $\mathcal P_N=\{(l,l'):1\le l<l'\le L,\ \sum_{i=n+1}^{n+N} m_i^{(l)}m_i^{(l')}>0\}$ and $\ell_{\mathrm{align}}$ can be the squared Euclidean distance or a contrastive alignment loss. We next regularize the empirical second moment of the concatenated representation toward the identity, which fixes its scale and discourages linear dependence both within and between the shared and modality-specific blocks. Define $W^{(l)}(\mathbf{x}_i^{(l)})=\left[U^{(l)}(\mathbf{x}_i^{(l)})^\top,V^{(l)}(\mathbf{x}_i^{(l)})^\top\right]^\top$ and let $\|\cdot\|_{\rm F}$ denote the Frobenius norm. We impose:
\begin{equation}
    \mathcal{L}_{\mathrm{orth}} = \frac{1}{L}\sum_{l=1}^L \left\| \frac{1}{\sum_{i=n+1}^{n+N} m_i^{(l)}} \sum_{i=n+1}^{n+N} m_i^{(l)} W^{(l)}(\mathbf{x}_i^{(l)}) W^{(l)}(\mathbf{x}_i^{(l)})^\top - \mathbf{I}_{d^{(0)}+d^{(l)}} \right\|_{\rm F}^2. 
    \label{eq:rep_orth_loss}
\end{equation}
However, orthogonality alone does not ensure that the modality-specific representation captures meaningful modality-specific information; it may instead learn arbitrary orthogonal noise. To prevent this, the representations are also required to retain information needed to reconstruct the original modality. With a decoder $T^{(l)}$, we use:
\begin{equation}
    \mathcal{L}_{\mathrm{recon}} = \frac{1}{L}\sum_{l=1}^L \frac{1}{\sum_{i=n+1}^{n+N} m_i^{(l)}}\sum_{i=n+1}^{n+N} m_i^{(l)} \left\| \mathbf{x}_{i}^{(l)} - T^{(l)}\left( \left[U^{(l)}(\mathbf{x}_i^{(l)}); V^{(l)}(\mathbf{x}_i^{(l)})\right] \right) \right\|_2^2.
    \label{eq:rep_recon_loss}
\end{equation}
Combining the three terms in \eqref{eq:rep_shared_loss}--\eqref{eq:rep_recon_loss}, the representation-learning objective is:
\begin{equation}
    \mathcal{L}_{\mathrm{total}}=\mathcal{L}_{\mathrm{shared}}+\lambda_o\mathcal{L}_{\mathrm{orth}}+\lambda_r\mathcal{L}_{\mathrm{recon}},
    \label{eq:full_rep_loss}
\end{equation}
where $\lambda_o,\lambda_r>0$ control the strengths of the last two terms. The shared loss places the outputs of the modality-specific shared encoders in a common coordinate system. The resulting shared and modality-specific representations determine which parts of a focal-pattern representation each observed pattern can supply.

\subsection{Preliminary Prediction from Available Representations}\label{sec:downstream_method}

For observations with pattern $A_j$, let $\mathbf Z_{(j)}=(\mathbf z_{1,(j)},\ldots,\mathbf z_{n_j,(j)})^\top\in\mathbb R^{n_j\times d_j}$ stack the representations indexed by $S_j$, and let $\mathbf Y_{(j)}\in\mathbb R^{n_j}$ be the corresponding response vector. MOSAIC uses all patterns to construct a different preliminary predictor for each focal pattern. For focal pattern $A_k$ and any observed pattern $A_j$, define $\mathbf Z_{(j\to k)}\in\mathbb R^{n_j\times d_k}$ by retaining the components indexed by $S_j\cap S_k$ and zero-filling those indexed by $S_k\setminus S_j$. Assuming that an observed modality-specific representation is nonzero almost surely, the zero-filled blocks distinguish unavailable representations from observed ones. This maps observations from every pattern into the focal-pattern input space $\mathbb R^{d_k}$, while retaining only the representation blocks supplied by their observed pattern.

To describe the focal-pattern information unavailable under $A_j$, define the complementary oracle matrix $\mathbf Z_{(-j\to k)}\in\mathbb R^{n_j\times d_k}$ by retaining the components indexed by $S_k\setminus S_j$ and zero-filling those indexed by $S_j\cap S_k$. Although these blocks are unobserved under $A_j$, they are well-defined in the latent representation model. Thus, $\mathbf Z_{(j\to k)}$ and $\mathbf Z_{(-j\to k)}$ partition the complete focal-pattern representation into blocks supplied and not supplied by $A_j$. For example, if $S_k=\{0,2,4\}$ and $S_j=\{0,2\}$, then $\mathbf z_{i,(j\to k)}=\left(\mathbf z_i^{(0)\top},\mathbf z_i^{(2)\top},\mathbf 0_{d^{(4)}}^\top\right)^\top$ and $\mathbf z_{i,(-j\to k)}=\left(\mathbf 0_{d^{(0)}}^\top,\mathbf 0_{d^{(2)}}^\top,\mathbf z_i^{(4)\top}\right)^\top$. Only $\mathbf Z_{(j\to k)}$ enters the estimator; the complementary matrix is used in the population analysis to characterize the effect of incomplete representation coverage. Let $\mathcal H_k$ be the first-stage function class. MOSAIC estimates the preliminary predictor by:
\begin{equation}
    \widehat h_k\in\argmin_{h\in\mathcal H_k} \frac{1}{n}\sum_{j=1}^K\left\|\mathbf Y_{(j)}-h(\mathbf Z_{(j\to k)})\right\|_2^2.
    \label{eq:first_step_estimator_general}
\end{equation}
Let $\mathbf z_{(j\to k)}$ denote a generic row of $\mathbf Z_{(j\to k)}$. Conditioning on the realized pattern counts, the corresponding population preliminary predictor is:
\[
    h_k:=\argmin_{h\in\mathcal{H}_k}\sum_{j=1}^K\frac{n_j}{n}\mathbb E\left[\{Y-h(\mathbf z_{(j\to k)})\}^2\mid \mathcal A=A_j\right].
\]

\subsection{Focal-Pattern Calibration}\label{sec:target_calibration}

The population preliminary predictor $h_k$ need not equal $g_{(k)}$ because other patterns can omit focal-pattern representations or have different conditional response relations. Calibration estimates the resulting correction $\Delta_k$ from observations with pattern $A_k$, where every representation in $S_k$ is available.

The statistical gain occurs when the representations required by $\widehat h_k$ are supplied by enough observations across patterns and $\Delta_k$ is easier to estimate from the focal-pattern observations than the full rule. Let $\mathcal D_k$ be a class of calibration functions and let $P(\cdot)$ be a penalty. MOSAIC estimates the correction from focal-pattern observations:
\begin{equation}
    \widehat\Delta_k\in\argmin_{\Delta\in\mathcal D_k}
    \frac{1}{n_k}\left\|\mathbf Y_{(k)}-\widehat h_k(\mathbf Z_{(k)})-\Delta(\mathbf Z_{(k)})\right\|_2^2
    +\lambda_k P(\Delta),\quad
    \widehat g_{(k)}:=\widehat h_k+\widehat\Delta_k.
    \label{eq:overlap_second_step_general}
\end{equation}
Here, $P(\Delta)$ controls the correction complexity when $n_k$ is small. Algorithm~\ref{alg:computation} summarizes the procedure.

\begin{algorithm}[htpb]
\caption{MOSAIC estimation procedure}
\label{alg:computation}
\begin{algorithmic}[1]
\Require unlabelled observations $\{(\mathcal X_i(\mathcal A_i),\mathcal A_i):i=n+1,\dots,n+N\}$ and labelled observations $\{(\mathcal X_i(\mathcal A_i),y_i,\mathcal A_i):i=1,\dots,n\}$; tuning parameters $\lambda_o,\lambda_r,\{\lambda_k\}_{k=1}^K$.

\State Estimate $\{\widehat U^{(l)}\}_{l=1}^L$ and $\{\widehat V^{(l)}\}_{l=1}^L$ by minimizing $\mathcal{L}_{\mathrm{total}}$.
\For{$i=1,\dots,n$}
    \State Compute $\widehat{\mathbf z}_i^{(0,l)}=\widehat U^{(l)}(\mathbf x_i^{(l)})$ and $\widehat{\mathbf z}_i^{(l)}=\widehat V^{(l)}(\mathbf x_i^{(l)})$ for every $l$ with $m_i^{(l)}=1$.
    \State Obtain the fused shared representation, for example by $\widehat{\mathbf z}_i^{(0)}=\frac{\sum_{l=1}^L m_i^{(l)}\widehat{\mathbf z}_i^{(0,l)}}{\sum_{l=1}^L m_i^{(l)}}$.
\EndFor

\State Construct $\mathbf Z_{(k)}$ and $\mathbf Y_{(k)}$ for each focal pattern $A_k$.
\For{$k=1,\dots,K$}
    \State Construct $\mathbf Z_{(j\to k)}$ for every $j=1,\dots,K$.
    \State Estimate the preliminary predictor $\widehat h_k$ by \eqref{eq:first_step_estimator_general}.
    \State Estimate the focal-pattern correction $\widehat\Delta_k$ by \eqref{eq:overlap_second_step_general}.
\EndFor

\State \Return $\{\widehat g_{(k)}=\widehat h_k+\widehat\Delta_k:k=1,\dots,K\}$.
\end{algorithmic}
\end{algorithm}

\subsection{Linear Model Specialization}\label{subsec:linear model}

We specialize the framework to linear prediction under the true complete-information model $Y=\bm\beta^{\star,\top}\mathbf z^{\star}+\varepsilon$, with $\mathbb E(\varepsilon\mid\mathbf z^{\star},\mathcal A)=0$, where $\bm\beta^{\star}\in\mathbb R^d$ is the complete-information coefficient. For pattern $A_k$, partition $\mathbf z^{\star}$ into the available representation $\mathbf z_{(k)}^{\star}$ indexed by $S_k$ and the complementary representation $\mathbf z_{(-k)}^{\star}$ indexed by $S_k^c=(\{0\}\cup[L])\setminus S_k$. The linear specialization of \eqref{eq:pattern_specific_target} is:
\[
    \bm\beta_{(k)}^{\star}:=\argmin_{\bm\beta\in\mathbb R^{d_k}}
    \mathbb E\left[\{Y-\bm\beta^{\top}\mathbf z_{(k)}^{\star}\}^2\mid\mathcal A=A_k\right],
    \qquad
    g_{(k)}(\mathbf z_{(k)}^{\star})=\bm\beta_{(k)}^{\star\top}\mathbf z_{(k)}^{\star}.
\]

Although $\bm\beta^{\star}$ is common, $\bm\beta_{(k)}^{\star}$ need not equal $\bm\beta_{S_k}^{\star}$ because the unavailable representation $\mathbf z_{(-k)}^{\star}$ may be associated with $\mathbf z_{(k)}^{\star}$. The following scalar two-pattern example shows how an unavailable but correlated representation changes a pattern-specific coefficient and why a pooled model that imposes a common coefficient on $z^{(1)}$ cannot generally recover both pattern-specific rules.

\begin{proposition}[Mismatch under indicator-augmented pooling]
\label{prop:linear-pooled-mismatch}
Let $d^{(0)}=d^{(1)}=d^{(2)}=1$, write $\bm\beta^{\star}=(\beta_0^{\star},\beta_1^{\star},\beta_2^{\star})^\top$, and suppose that $z^{\star,(0)},z^{\star,(1)}$, and $z^{\star,(2)}$ are mean-zero and unit-variance, with $z^{\star,(0)}\perp (z^{\star,(1)},z^{\star,(2)})$ and $\operatorname{Cov}(z^{\star,(1)},z^{\star,(2)})=\rho$. Consider two observed-modality patterns $A_1=\{1\}$ and $A_2=\{1,2\}$, with representation index sets $S_1=\{0,1\}$ and $S_2=\{0,1,2\}$. Let $m^{(2)}=\mathbf 1\{\mathcal A=A_2\}\sim\mathrm{Bernoulli}(\pi)$ be independent of $(z^{\star,(0)},z^{\star,(1)},z^{\star,(2)},\varepsilon)$. Then the pattern-specific linear coefficients are $\bm\beta_{(1)}^{\star}=(\beta_0^{\star},\beta_1^{\star}+\rho\beta_2^{\star})^\top$ and $\bm\beta_{(2)}^{\star}=(\beta_0^{\star},\beta_1^{\star},\beta_2^{\star})^\top$, whereas the population coefficient from the pooled design $(z^{\star,(0)},z^{\star,(1)},m^{(2)},m^{(2)}z^{\star,(2)})^\top$ is
\[
    \bm{\beta}_{\mathrm{pool}}^{\star}=\left(\beta_0^{\star},\beta_1^{\star}+\frac{(1-\pi)\rho\beta_2^{\star}}{1-\pi\rho^2},0,\frac{(1-\rho^2)\beta_2^{\star}}{1-\pi\rho^2}\right)^\top.
\]
Except in degenerate cases such as $\rho\beta_2^{\star}=0$, the rules induced by $\bm\beta_{\mathrm{pool}}^{\star}$ do not coincide with either pattern-specific rule.
\end{proposition}
The pooled model in Proposition~\ref{prop:linear-pooled-mismatch} constrains the coefficient of $z^{\star,(1)}$ to be the same in both patterns. This restriction can be relaxed by allowing covariate coefficients to depend on the missingness pattern, for example through covariate--indicator interactions \citep{mercaldo2020missing, bertsimas2025adaptive}. In the present example, adding $m^{(2)}z^{\star,(1)}$ recovers both pattern-specific rules at the population level. However, increasing such pattern indicator introduces additional pattern-dependent parameters. It reduces to separate pattern-specific fitting \citep{mercaldo2020missing, bertsimas2025adaptive}, which can be inefficient when some pattern samples are small. This motivates borrowing across patterns while retaining target-pattern calibration rather than imposing either common or fully pattern-specific coefficients.

We now return to the general $K$-pattern model and specialize the two stages of MOSAIC. For a focal pattern $A_k$ and every observed pattern $A_j$, write in matrix form:
\[
    \mathbf Y_{(j)}
    =\bigl(\mathbf Z_{(j\to k)}^{\star}+\mathbf Z_{(-j\to k)}^{\star}\bigr)\bm\beta^{\star}_{(k)}
    +\mathbf r^{\star}_{(j,k)},\quad j=1,\ldots,K,
\]
where $\mathbf r_{(j,k)}^{\star}$ is the residual from applying the focal-pattern rule to observations with pattern $A_j$. If $\mathcal H_k=\{h_{\bm w}:h_{\bm w}(\mathbf Z)=\mathbf Z\bm w,\bm w\in\mathbb R^{d_k}\}$, then \eqref{eq:first_step_estimator_general} becomes:
\[
    \widehat{\bm w}_{(k)}^\star\in\argmin_{\bm w\in\mathbb R^{d_k}}\frac{1}{n}\sum_{j=1}^K\left\|\mathbf Y_{(j)}-\mathbf Z_{(j\to k)}^{\star}\bm w\right\|_2^2,\quad\widehat h_k(\mathbf Z)=\mathbf Z\widehat{\bm w}_{(k)}^{\star}.
\]
Its population coefficient is $\bm w_{(k)}^{\star}:=\argmin_{\bm w\in\mathbb R^{d_k}}\sum_{j=1}^K\frac{n_j}{n}\mathbb E\left[\{Y-\mathbf z_{(j\to k)}^{\star,\top}\bm w\}^2\mid \mathcal A=A_j\right]$. Define the corresponding weighted moment matrices as:
\[
    \widetilde{\mathbf\Sigma}_{(k)}^{oo,\star}=\sum_{j=1}^K\frac{n_j}{n}\widetilde{\mathbf\Sigma}_{j,k}^{oo,\star},\quad\widetilde{\mathbf\Sigma}_{(k)}^{om,\star}=\sum_{j=1}^K\frac{n_j}{n}\widetilde{\mathbf\Sigma}_{j,k}^{om,\star},
\]
where $\widetilde{\mathbf\Sigma}_{j,k}^{oo,\star}=\mathbb E[\mathbf z_{(j\to k)}^{\star}\mathbf z_{(j\to k)}^{\star,\top}\mid \mathcal A=A_j], \widetilde{\mathbf\Sigma}_{j,k}^{om,\star}=\mathbb E[\mathbf z_{(j\to k)}^{\star}\mathbf z_{(-j\to k)}^{\star,\top}\mid \mathcal A=A_j]$.
Under the cross-pattern moment condition $\mathbb E[\mathbf z_{(j\to k)}^{\star}r_{(j,k)}^{\star}\mid\mathcal A=A_j]=\mathbf 0$ for every $j\in[K]$, the normal equations give $\bm w_{(k)}^{\star}=\bm\beta_{(k)}^{\star}+\bigl(\widetilde{\mathbf\Sigma}_{(k)}^{oo,\star}\bigr)^{-1}\widetilde{\mathbf\Sigma}_{(k)}^{om,\star}\bm\beta_{(k)}^{\star}$.
Thus, the coefficient of the preliminary predictor can differ from $\bm\beta_{(k)}^{\star}$. Write the coefficient correction as $\bm\delta_{(k)}^{\star}:=\bm\beta_{(k)}^{\star}-\bm w^{\star}_{(k)}$ and define $\widetilde\rho^{\star}_{(k)}=\|(\widetilde{\mathbf\Sigma}_{(k)}^{oo,\star})^{-1}\widetilde{\mathbf\Sigma}_{(k)}^{om,\star}\|_{\mathrm{op}}$. This quantity measures how strongly the representations supplied across patterns are associated with focal-pattern representations omitted in some patterns, and $\|\bm\delta_{(k)}^{\star}\|_2\le \widetilde\rho^{\star}_{(k)}\|\bm\beta_{(k)}^{\star}\|_2$. If $\mathcal D_k=\{\Delta_{\bm\delta}:\Delta_{\bm\delta}(\mathbf Z)=\mathbf Z\bm\delta,\bm\delta\in\mathbb R^{d_k}\}$ and $P(\Delta)=\|\bm\delta\|_2$, the calibration estimator becomes:
\[
    \widehat{\bm\delta}_{(k)}^{\star}\in\argmin_{\bm\delta\in\mathbb R^{d_k}}
    \frac{1}{n_k}\left\|\mathbf Y_{(k)}-\mathbf Z_{(k)}^{\star}
    \bigl(\widehat{\bm w}_{(k)}^{\star}+\bm\delta\bigr)\right\|_2^2
    +\lambda_k\|\bm\delta\|_2,
    \qquad
    \widehat{\bm\beta}_{(k)}^{\star}
    :=\widehat{\bm w}_{(k)}^{\star}+\widehat{\bm\delta}_{(k)}^{\star}.
\]

\section{Theory}\label{sec theory}
The theory below specializes MOSAIC to the linear model in Section~\ref{subsec:linear model}. We first study cross-pattern borrowing and focal-pattern calibration under oracle representations. We then establish recovery of the representations and finally propagate representation error into prediction. We first distinguish the true representations $\mathbf{z}^{\star,(l)}_i$ from their learned counterparts $\widehat{\mathbf{z}}^{(l)}_i$ and then introduce the assumptions required for the theoretical analysis.

\begin{assumption}[Missing Completely at Random (MCAR)]\label{ass:missing}
    Assume that $\mathbf{m}_i=(m_i^{(1)},\dots,m_i^{(L)})^\top$ is independent of $(\mathbf z_i^\star,\bm\varepsilon_{x,i},Y_i)$. For every $l,l^\prime\in [L]$ and missingness pattern $k\in[K]$, $\mathrm{Pr}(m_i^{(l)}=1, m_i^{(l^\prime)}=1)= q^{(ll^\prime)}\ge \underline{q}_N>0$ and $\pi_k\ge \underline{q}_{n,k}>0$, where $\underline{q}_N\asymp N^{-\gamma}, \underline{q}_{n,k}\asymp n^{-\iota_k}, \gamma,\iota_k\in[0,1)$. Let $\mathcal{H}^{s}=\{k\in[K]: |A_k|=1\}$. We further assume that there exists a constant $c_{\iota}>0$ such that $\min_{k\in [K]\setminus\mathcal{H}^{s}}\iota_k-2\max_{k\in\mathcal{H}^{s}}\iota_k\ge c_{\iota}$.
\end{assumption}

\begin{assumption}[Prediction Noise]\label{ass:noise for y}
    For each $k,j\in[K]$: (i) $\mathbb E[Y-\bm\beta_{(k)}^{\star\top}\mathbf z_{(k)}^\star \mid\mathcal A=A_k]=0$; (ii) $\mathbb{E}\{\mathbf{z}_{(j \to k),i}^{\star} r_{(j, k),i}^{\star}\mid\mathcal A=A_j\}=\bm{0}$; (iii) For any $t\in\mathbb{R}$, $\mathbb{E}\{\exp(tr_{(j, k),i}^{\star})\mid\mathcal A=A_j\}\le\exp(\tfrac12\tau^2 t^2)$.
\end{assumption}

\begin{assumption}[Representation]\label{ass:representation}
    For each representation $l\in \{0\}\cup[L]$: (i) $\mathbb{E}[\mathbf{z}_i^{\star,(l)}]=\mathbf{0}$; (ii) $\operatorname{Cov}(\mathbf{z}_i^{\star,(l)})=\mathbf{I}_{d^{(l)}}$; (iii) For any unit $\mathbf{u}\in\mathbb{R}^{d^{(l)}}$ and $t\in\mathbb{R}$, $\mathbb{E}\!\left[\exp\{t\langle\mathbf{u},\mathbf{z}_i^{\star,(l)}\rangle\}\right]\le\exp(\tfrac12\tau^2 t^2)$; (iv) $\operatorname{Cov}(\mathbf{z}_i^{\star,(l)},\mathbf{z}_i^{\star,(l')})=\mathbf{0}$ for $l\neq l'$.
\end{assumption}

For the theoretical analysis, Assumption~\ref{ass:missing} allows the relevant observation and pattern probabilities to decay with sample size under MCAR controlled by $\gamma$ and $\iota_k$. The final condition imposes a separation between the rarity orders of single-modality and multiple-modalities patterns; its implication for cross-pattern borrowing is discussed after Theorem~\ref{thm:main_transfer}.
Assumption~\ref{ass:noise for y} provides the moment and tail conditions for the noise, and Assumption~\ref{ass:representation} controls the covariance structure and sub-Gaussian tails of the true representations. The unit-covariance normalization fixes their scale. 

Let $n_{k,(l)}=\sum_{j:\,l\in S_j}n_j$ be the number of first-stage observations supplying focal-pattern representation $l\in S_k$. Define the effective sample size as $\widetilde n_{(k)}:=n\left\{\frac{\min_{l\in S_k}n_{k,(l)}}{n}\right\}^2$. We are now ready to establish the oracle prediction error bound.
\begin{theorem}\label{thm:main_transfer}
    Fix a focal pattern $A_k$ and condition on the realized missingness patterns. Suppose that $n_k\gtrsim d_k+\log n$ and $\widetilde n_{(k)}\gtrsim d_k+\log n$. Under Assumptions~\ref{ass:missing}, \ref{ass:noise for y}, and \ref{ass:representation}(i)--(iii), if $\lambda_k\asymp\sqrt{(d_k+\log n)/n_k}$, then, with probability at least $1-Cn^{-1}$,
    \begin{equation}
      \|\widehat{\bm\beta}_{(k)}^\star-\bm\beta_{(k)}^\star\|_2^2\lesssim\bigl\{1+(1+\widetilde\rho_{(k)}^\star)\|\bm\beta_{(k)}^\star\|_2\bigr\}^2\frac{d_k+\log n}{\widetilde n_{(k)}}+\min\left\{\frac{d_k+\log n}{n_k},\|\bm\delta_{(k)}^\star\|_2^2\right\}.
      \label{eq:main_transfer_bound_main}
    \end{equation}
    If Assumption~\ref{ass:representation}(iv) also holds, $\|\bm\beta_{(k)}^\star\|_2$ is uniformly bounded, and the multiplicative constant in $\lambda_k$ is sufficiently large, then simultaneously for all $l\in S_k$,
    \begin{equation}
        \left\|\widehat{\bm\beta}_{(k)}^{\star,(l)}-\bm\beta_{(k)}^{\star,(l)}\right\|_2^2\lesssim\bigl(1+\|\bm\beta_{(k)}^\star\|_2\bigr)^2\frac{d_k+\log n}{n_{k,(l)}}.
        \label{eq:componentwise-oracle-rate}
    \end{equation}
    In particular, since $n_{k,(0)}=n$, $\left\|\widehat{\bm\beta}_{(k)}^{\star,(0)}-\bm\beta_{(k)}^{\star,(0)}\right\|_2^2\lesssim\bigl(1+\|\bm\beta_{(k)}^\star\|_2\bigr)^2\frac{d_k+\log n}{n}$.
\end{theorem}
The first term in \eqref{eq:main_transfer_bound_main} is the preliminary-predictor estimation error and is governed by the focal-pattern representation supplied by the fewest observations. The second term is the calibration error: a small correction is inexpensive, while a large correction returns this component to the local scale. Thus, the general bound makes explicit how representation coverage and focal-pattern correction jointly determine the gain from cross-pattern borrowing. Under Assumption~\ref{ass:representation}(iv), the population correction vanishes, and a sufficiently large calibration penalty also removes the empirical correction. The componentwise bound in \eqref{eq:componentwise-oracle-rate} then retains the observation count for each representation block rather than reducing all blocks to the least-covered representation.

\change{For the shared representation, $n_{k,(0)}=n$. Thus, when $n_k=o(n)$, the shared coefficient has a strictly faster rate under cross-pattern borrowing.} This all-pattern gain can include observations with no raw covariate in common with the focal pattern: if $A_j\cap A_k=\varnothing$, then $S_j\cap S_k=\{0\}$, so the $n_j$ observations with pattern $A_j$ contribute to $n_{k,(0)}$. 
\change{For comparison, we modify the linear model in Section~\ref{subsec:linear model} by replacing the latent representations with the corresponding raw modalities. For each raw modality $l$, we decompose its coefficient under missingness pattern $k$ as a common coefficient vector $\bm\theta^{\star,(l)}$ plus a sparse pattern-specific deviation $\bm\Delta_k^{\star,(l)}$, for $l\in A_k$.} This formulation also covers SPSM \citep{stempfle2023sharing}. \change{Proposition~\ref{prop:spsm-mosaic-borrowing} then provides a theoretical result under this raw-modality formulation.}
\begin{proposition}[Model without representations]\label{prop:spsm-mosaic-borrowing}
    Fix a focal pattern $A_k$. For each $l\in A_k$, write $\bm\theta^{\star,(l)}+\bm\Delta_k^{\star,(l)}$ for the coefficient of raw modality $l$ under $A_k$, and let $s_k^{(l)}=\|\bm\Delta_k^{\star,(l)}\|_0$. Suppose that the model is correctly specified. Under the regularity conditions stated in Supplementary Section~A.5, with probability at least $1-Cn^{-1}$, simultaneously for all $l\in A_k$,
    \begin{equation}
        \left\|(\widehat{\bm\theta}^{(l)}+\widehat{\bm\Delta}_k^{(l)})-(\bm\theta^{\star,(l)}+\bm\Delta_k^{\star,(l)})\right\|_2^2\lesssim\frac{d_{\rm raw}^{(l)}+\log n}{n_{k,(l)}}+\min\left\{\frac{s_k^{(l)}\log\{n d_{\rm raw}^{(l)}\}}{n_k},\|\bm\Delta_k^{\star,(l)}\|_2^2\right\}.
    \end{equation}
\end{proposition}
Proposition~\ref{prop:spsm-mosaic-borrowing} and Theorem~\ref{thm:main_transfer} show two further gains from MOSAIC, beyond $n_{k,(0)}=n$. 
First, representation replaces the raw covariate dimension by the much smaller focal-pattern representation dimension $d_k$, yielding a lower-dimensional estimation problem whenever $d_k\ll d_{\rm raw}^{(l)}$.  
Second, if we do not disentangle the raw covariates into shared and specific representations, orthogonality between the raw covariates is not guaranteed. In contrast, we impose Assumption~\ref{ass:representation}(iv), allowing the second term in the error bound to vanish.

In addition, the separation condition in Assumption~\ref{ass:missing} yields a strict improvement over
focal-only fitting, by describing a setting in which patterns containing multiple modalities are less frequent than single-modality patterns. If every $l\in A_k$ has a corresponding single-modality pattern $h^{(l)}$, then $n_{k,(l)}\ge n_{h^{(l)}}$ and, with high probability, $n_{k,(l)}\gtrsim n^{1-\max_{j\in\mathcal H^s}\iota_j}$ for every $l\in A_k$. If the lower envelope is also sharp for a multi-modality focal pattern, so that $\pi_k\asymp n^{-\iota_k}$ and hence $n_k\asymp n^{1-\iota_k}$ with high probability, then $n_{k,(l)}/n_k\gtrsim n^{\iota_k-\max_{j\in\mathcal H^s}\iota_j}\gtrsim n^{c_\iota}\to\infty$. Therefore, under \eqref{eq:componentwise-oracle-rate}, every focal modality-specific coefficient block has an error of $o\{(d_k+\log n)/n_k\}$, and the shared coefficient has the all-pattern rate based on $n$. In the general case without Assumption~\ref{ass:representation}(iv), the same count comparison gives $\widetilde n_{(k)}\gtrsim n^{1-2\max_{j\in\mathcal H^s}\iota_j}$ and $\widetilde n_{(k)}/n_k\gtrsim n^{c_\iota}\to\infty$. When $\{1+(1+\widetilde\rho_{(k)}^\star)\|\bm\beta_{(k)}^\star\|_2\}^2=O(1)$ and $\|\bm\delta_{(k)}^\star\|_2^2=o\{(d_k+\log n)/n_k\}$, \eqref{eq:main_transfer_bound_main} then gives an overall error of $o\{(d_k+\log n)/n_k\}$.

Theorem~\ref{thm:main_transfer} is conditional on the realized pattern counts. Supplementary Section~B.2 shows that replacing the realized proportions $n_j/n$ by the population prevalences $\pi_j$ introduces only a perturbation; for fixed $K$, the additional squared-error term is of order $\|\bm\beta_{(k)}^\star\|_2^2\log n/n$.

We next study recovery of the oracle representations used in Theorem~\ref{thm:main_transfer}. For notational simplicity, we re-index the unlabelled sample as $i=1,\ldots,N$; the independent labelled sample continues to have size $n$. We define $\mathbf{x}_i^{(l)}\in\mathbb{R}^{d_{\rm{raw}}^{(l)}}$ using \eqref{eq:gen_model_x} with fixed matrices $\mathbf{A}^{(l)} \in\mathbb{R}^{d_{\rm{raw}}^{(l)} \times d^{(0)}}, \mathbf{B}^{(l)} \in\mathbb{R}^{d_{\rm{raw}}^{(l)} \times d^{(l)}}$, and $\mathbf{C}^{(l)} \in\mathbb{R}^{d_{\rm{raw}}^{(l)} \times d_{\rm{noise}}^{(l)}}$:
\begin{equation}
    \mathbf{x}_i^{(l)} = \mathbf{A}^{(l)} \mathbf{z}_i^{\star, (0)} + \mathbf{B}^{(l)} \mathbf{z}_i^{\star, (l)} + \mathbf{C}^{(l)} \bm{\varepsilon}_{x,i}^{(l)},
    \label{eq:gen_model_linear}
\end{equation}
where $d^{(l)}_{\rm{noise}}:=d_{\rm{raw}}^{(l)}-d^{(0)}-d^{(l)}$ and $\max(d^{(0)},d^{(l)})\ll d^{(l)}_{\rm{noise}}$. This high-dimensional ambient-noise regime is used only to analyse representation recovery; it is not required by Algorithm~\ref{alg:computation}. We model the corresponding encoders as matrices $\mathbf{U}^{(l)}$ and $\mathbf{V}^{(l)}$. The ideal encoders that recover the shared and modality-specific representations separately are $\mathbf{U}_{\rm{ideal}}^{(l)} = (\mathbf{A}^{(l)\top} \mathbf{A}^{(l)})^{-1} \mathbf{A}^{(l)\top}$ and $\mathbf{V}_{\rm{ideal}}^{(l)} = (\mathbf{B}^{(l)\top} \mathbf{B}^{(l)})^{-1} \mathbf{B}^{(l)\top}$. 

Let the joint encoder be defined as $\mathbf{W}^{(l)} = [\mathbf{U}^{(l)\top}, \mathbf{V}^{(l)\top}]^\top \in\mathbb{R}^{(d^{(0)}+d^{(l)}) \times d_{\rm{raw}}^{(l)}}$, and let $\mathbf{T}^{(l)}$ be the corresponding linear decoder matrix. Define the number of co-observed samples for modality pair $(l,k)$ as $N_{lk}=\sum_{i=1}^N m_i^{(l)}m_i^{(k)}$ and the aggregated empirical covariance of modality $l$ as $\widetilde{\mathbf{\Sigma}}_{ll}=L^{-1}\sum_{k=1}^L N_{lk}^{-1}\sum_{i=1}^N m_i^{(l)}m_i^{(k)}\mathbf{x}_i^{(l)}\mathbf{x}_i^{(l)\top}$. We work on the event that all $N_{lk}$ are positive; Assumption~\ref{ass:missing} and the stated sample-size conditions ensure that this event has high probability. For a tractable linear analysis, we study the following surrogate of \eqref{eq:full_rep_loss}:
\[
    \begin{aligned}
        &\min_{\{\mathbf{W}^{(l)},\mathbf{T}^{(l)}\}_{l=1}^L} \frac{1}{L^2} \sum_{l,k=1}^L \frac{1}{N_{lk}} \sum_{i=1}^N  m_i^{(l)}m_i^{(k)}\left(\| \mathbf{U}^{(l)} \mathbf{x}_{i}^{(l)} - \mathbf{U}^{(k)} \mathbf{x}_{i}^{(k)} \|_2^2 + \lambda\|\mathbf{T}^{(l)}\mathbf{W}^{(l)}\mathbf{x}_i^{(l)}-\mathbf{x}_i^{(l)}\|_2^2\right)\\
        &\qquad\text{s.t. } \mathbf{W}^{(l)}\widetilde{\mathbf{\Sigma}}_{ll}\mathbf{W}^{(l)\top} = \mathbf{I}_{d^{(0)}+d^{(l)}},
    \end{aligned}
\]
where $\lambda > 0$ is a tuning parameter. This surrogate replaces the soft orthogonalization penalty in \eqref{eq:full_rep_loss} by a hard whitening constraint and uses a single weight for reconstruction. The hard constraint can be imposed directly in the linear model; details are provided in Supplementary Section~A.2. We further impose the following assumptions.

\begin{assumption}[Structure]\label{ass:structure}
    For every modality $l$, $\mathbf{A}^{(l)}$ and $\mathbf{B}^{(l)}$ have full column rank, while $\mathbf{C}^{(l)}$ is allowed to be rank-deficient. Furthermore, $(\mathbf{A}^{(l)})^\top \mathbf{B}^{(l)} = \mathbf{0}$, $(\mathbf{A}^{(l)})^\top \mathbf{C}^{(l)} = \mathbf{0}$, and $(\mathbf{B}^{(l)})^\top \mathbf{C}^{(l)} = \mathbf{0}$. Let $\sigma_i(\mathbf M)$ denote the $i$-th singular value of a matrix $\mathbf M$. There exist positive constants $0 < \nu \le \mu$ such that every singular value of $\mathbf{A}^{(l)}$ and $\mathbf{B}^{(l)}$ satisfies $\sigma_i(\mathbf{A}^{(l)}), \sigma_i(\mathbf{B}^{(l)}) \in [\nu, \mu]$, and every singular value of $\mathbf{C}^{(l)}$ satisfies $\sigma_i(\mathbf{C}^{(l)}) \leq \frac{1}{\sqrt{2}}\nu$.
\end{assumption}

\begin{assumption}[Representation Noise]\label{ass:noise for x}
    For each modality $l \in [L]$: (i) $\mathbb{E}[\bm{\varepsilon}_{x,i}^{(l)}]=\mathbf{0}$; (ii) $\operatorname{Cov}(\bm{\varepsilon}_{x,i}^{(l)},\mathbf{z}_i^{\star,(k)})=\mathbf{0}$ for $k\in\{0,\ldots,L\}$; (iii) $\operatorname{Cov}(\bm{\varepsilon}_{x,i}^{(l)})=\mathbf{I}_{d_{\rm{noise}}^{(l)}}$; (iv) For any unit $\mathbf{u}\in\mathbb{R}^{d^{(l)}_{\rm{noise}}}$ and $t\in\mathbb{R}$, $\mathbb{E}\!\left[\exp\{t\langle\mathbf{u},\bm{\varepsilon}_{x,i}^{(l)}\rangle\}\right]\le\exp(\tfrac12\tau^2 t^2)$; (v) $\operatorname{Cov}(\bm{\varepsilon}_{x,i}^{(l)},\bm{\varepsilon}_{x,i}^{(k)})=\mathbf 0$ for $l\neq k$.
\end{assumption}

Assumption~\ref{ass:structure} imposes geometric separation of the shared and modality-specific representation subspaces, which makes the representations identifiable up to an orthogonal transformation in the linear analysis.
Assumption~\ref{ass:noise for x} controls the sub-Gaussian tails of the modality-level noise.
Under Assumptions~\ref{ass:representation}--\ref{ass:noise for x}, the shared representation is identifiable up to one common orthogonal transformation across modalities, while each modality-specific representation is identifiable up to its own orthogonal transformation; the proof is given in Supplementary Section~A.2. Let $d_{\max}=\max_{l\in[L]}d_{\rm{raw}}^{(l)}$ and let $\mathcal O(r)$ denote the set of $r\times r$ orthogonal matrices. The following theorem bounds the distances between the estimated encoders and their ideal counterparts, up to these transformations.

\begin{theorem}\label{thm:Representation Theory}
Under Assumption~\ref{ass:missing} and Assumptions~\ref{ass:representation}--\ref{ass:noise for x}, suppose the sample size $N$ is sufficiently large such that $\epsilon_{N,d_{\max}} \leq \frac{1}{8}\nu^2$, and the regularization parameter satisfies $\lambda\leq \min(\epsilon_{N,d_{\max}},\frac{L-2}{L\mu^2})$ with $L\geq 3$, where $\epsilon_{N,d_{\max}}=C_1(d_{\max}+\log N)^{1/2}N^{-(1-\gamma)/2}$ is detailed in the proof. Then, with probability at least $1-c\exp(-c^{\prime}N^{1-\gamma})-CN^{-1}$, the empirical encoders $(\widehat{\mathbf{U}}^{(l)}, \widehat{\mathbf{V}}^{(l)})$ converge to $(\mathbf{U}_{\rm{ideal}}^{(l)}, \mathbf{V}_{\rm{ideal}}^{(l)})$ up to the corresponding orthogonal transformations at the following rate:
\[
    \begin{aligned}
        \max\Big\{\min_{\mathbf{R}_0 \in \mathcal{O}(d^{(0)})}\frac{1}{L}\sum_{l=1}^L \| \widehat{\mathbf{U}}^{(l)} -  \mathbf{R}_0\mathbf{U}_{\rm{ideal}}^{(l)} \|_{\rm{F}},& \frac{1}{L}\sum_{l=1}^L\min_{\mathbf{R}_l \in \mathcal{O}(d^{(l)})} \| \widehat{\mathbf{V}}^{(l)} -  \mathbf{R}_l \mathbf{V}_{\rm{ideal}}^{(l)}\|_{\rm{F}}\Big\}\\
        &\lesssim \{d_{\max}(d_{\max}+\log N)\}^{1/2}N^{-(1-\gamma)/2}, 
    \end{aligned}
\]
where $C,c,c^{\prime}$ are positive constants independent of $(N,d_{\max})$.
\end{theorem}
Theorem~\ref{thm:Representation Theory} establishes that the shared and modality-specific representations are recoverable. Since $\gamma<1$, the representation error converges to zero as the unlabelled sample size $N$ increases. The restriction $L\ge3$ is used for spectral separation in this theorem; Algorithm~\ref{alg:computation} also applies to two modalities. We use the bound below to propagate representation-learning error into prediction. For a labelled subject, define $\widehat{\mathbf z}_i^{(0)}=\{\sum_{l=1}^Lm_i^{(l)}\}^{-1}\sum_{l=1}^Lm_i^{(l)}\widehat{\mathbf U}^{(l)}\mathbf x_i^{(l)}$ and $\widehat{\mathbf z}_i^{(l)}=\widehat{\mathbf V}^{(l)}\mathbf x_i^{(l)}$. The common unidentified rotation of the shared representation and the separate rotations of the modality-specific representations are absorbed into the corresponding oracle coordinate systems; rotating the associated coefficients in the same way leaves the prediction rule unchanged.

We next consider the learned-representation estimator used by MOSAIC. Under the orthogonality conditions in Assumption~\ref{ass:structure}, the ideal encoders recover the true representations exactly, satisfying $\mathbf U_{\rm{ideal}}^{(l)}\mathbf x_i^{(l)} = \mathbf z_i^{\star,(0)}$ and $\mathbf V_{\rm{ideal}}^{(l)}\mathbf x_i^{(l)} = \mathbf z_i^{\star,(l)}$. Together with Theorem~\ref{thm:Representation Theory}, this means that the learned design matrices can be treated as perturbations of the oracle design matrices in Theorem~\ref{thm:main_transfer}. Assumption~\ref{ass:representation} makes the population correction vanish in this orthogonal linear specialization, so the next theorem isolates the additional cost of learning the representations.

\begin{theorem}\label{thm:learned_transfer}
    Suppose the conditions of Theorem~\ref{thm:main_transfer} hold, except that the tuning parameter is specified below, and let $\widehat{\bm\beta}_{(k)}$ be the estimator constructed from the learned representations $\widehat{\bm{z}}^{(l)}_i$. Let $\eta_N=\{d_{\max}(d_{\max}+\log N)\}^{1/2}N^{-(1-\gamma)/2}$, and assume that the representation-learning and labelled samples are independent. Define $\bar\iota_k=\max_{l\in S_k}\min_{j:\,l\in S_j}\iota_j$, and suppose that the realized pattern counts satisfy $\max_{l\in S_k}n_{k,(l)}/\min_{l\in S_k}n_{k,(l)}\lesssim n^{\bar\iota_k}$.
    Suppose that $n^{\bar\iota_k}\eta_N+\sqrt{(d_k+\log N)/n}+\sqrt{(d_k+\log N)/n_k}$ is sufficiently small and that $\lambda_k\asymp\sqrt{(d_k+\log N)/n_k}+\eta_N$. Under Assumptions~\ref{ass:missing}--\ref{ass:noise for x}, with probability at least $1-c\exp(-c'N^{1-\gamma})-CN^{-1}$,
    \begin{equation}
      \|\widehat{\bm\beta}_{(k)}-\bm\beta_{(k)}^\star\|_2^2
      \lesssim
      \bigl(1+\|\bm\beta_{(k)}^\star\|_2\bigr)^2
      \left\{\frac{d_k+\log N}{\widetilde n_{(k)}}
      +n^{2\bar\iota_k}\eta_N^2\right\}.
      \label{eq:learned_transfer_bound_main}
    \end{equation}
\end{theorem}
The first term in Theorem~\ref{thm:learned_transfer} is the oracle prediction rate. The second is the representation-learning cost, amplified when the numbers of observations supplying different focal-pattern representations are imbalanced. When the second term is smaller, the learned-representation estimator has the same leading rate as its oracle counterpart. Supplementary Section~A.1 extends this result to generalized linear models, allowing representation-learning error and a potentially nonzero focal-pattern correction to enter the same bound; proofs are provided in Supplementary Appendix~D.

\section{Simulation Studies}
\label{sec:simulation}

The simulations are organized around the two stages of MOSAIC. Experiment~1 assesses recovery of the shared and modality-specific representations under missing modalities. Experiment~2 assesses downstream prediction after representation learning, with emphasis on whether overlap borrowing followed by focal-pattern correction improves performance across missingness patterns.

\subsection{Experiment 1: Representation Recovery}
\label{sec:simulation-exp1}

Experiment~1 examines the representation-learning stage by varying the sample size $N$ and the modality observation probability $q$. We also include ablations that remove orthogonal decoupling, reconstruction, or both, so that the roles of the representations can be separated. We generate $L=3$ modalities from a latent structure with shared and modality-specific representations. Each subject has a shared representation $\mathbf z_i^{\star,(0)}\in\mathbb R^5$, modality-specific representations $\mathbf z_i^{\star,(l)}\in\mathbb R^5$, and noise. We consider a linear anisotropic mixing regime and a nonlinear residual-tanh regime, with details given in Supplementary Section~E.1. For each subject and modality, $m_i^{(l)}\sim\mathrm{Bernoulli}(q)$ independently.

We compare MOSAIC with three ablations: MOSAIC (w/o orth), which removes the covariance-orthogonality penalty; MOSAIC (w/o recon), which removes reconstruction; and MOSAIC (w/o orth \& recon), which keeps only the shared alignment loss. We also include spectral SVD, a non-iterative baseline based on pairwise cross-modal covariance and residual within-modality variation, and the complete algorithm is given in Supplementary Section~E.1. For the sample-size sweep, we fix $q=0.6$ and vary $N$. For the missingness sweep, we fix $N=5000$ and vary $q$. Each condition uses 100 Monte Carlo replications. Recovery is measured by no-scale Procrustes error at the subject-embedding level after column centring, with details given in Supplementary Section~E.1. 

\begin{figure}[ht]
\centering
\includegraphics[width=\textwidth]{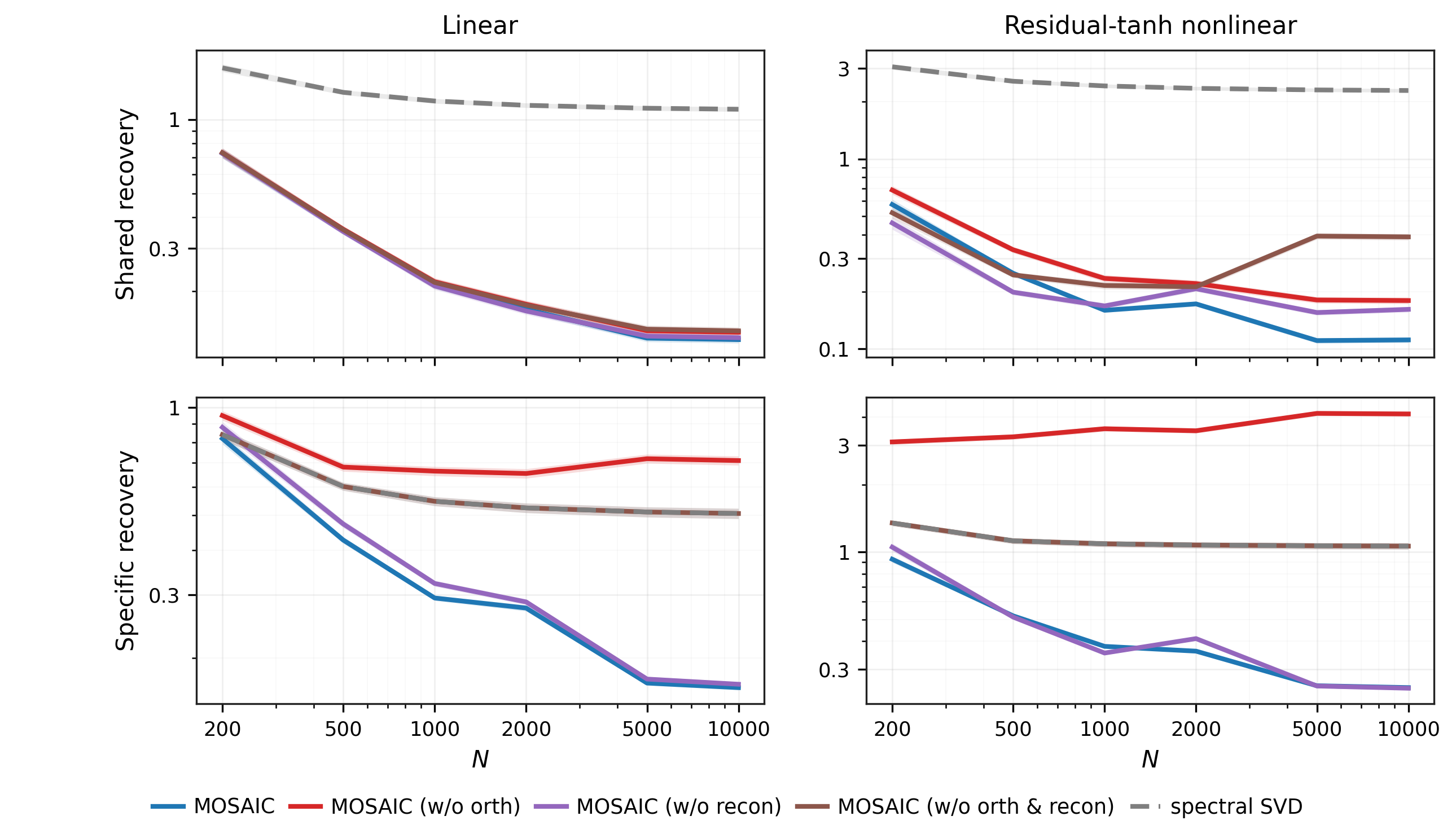}
\caption{Representation recovery in the Experiment~1 sample-size sweep. Shaded bands are pointwise 95\% Monte Carlo confidence intervals for the mean over 100 replications.}
\label{fig:exp1-sample-size-recovery}
\end{figure}

\begin{figure}[ht]
\centering
\includegraphics[width=\textwidth]{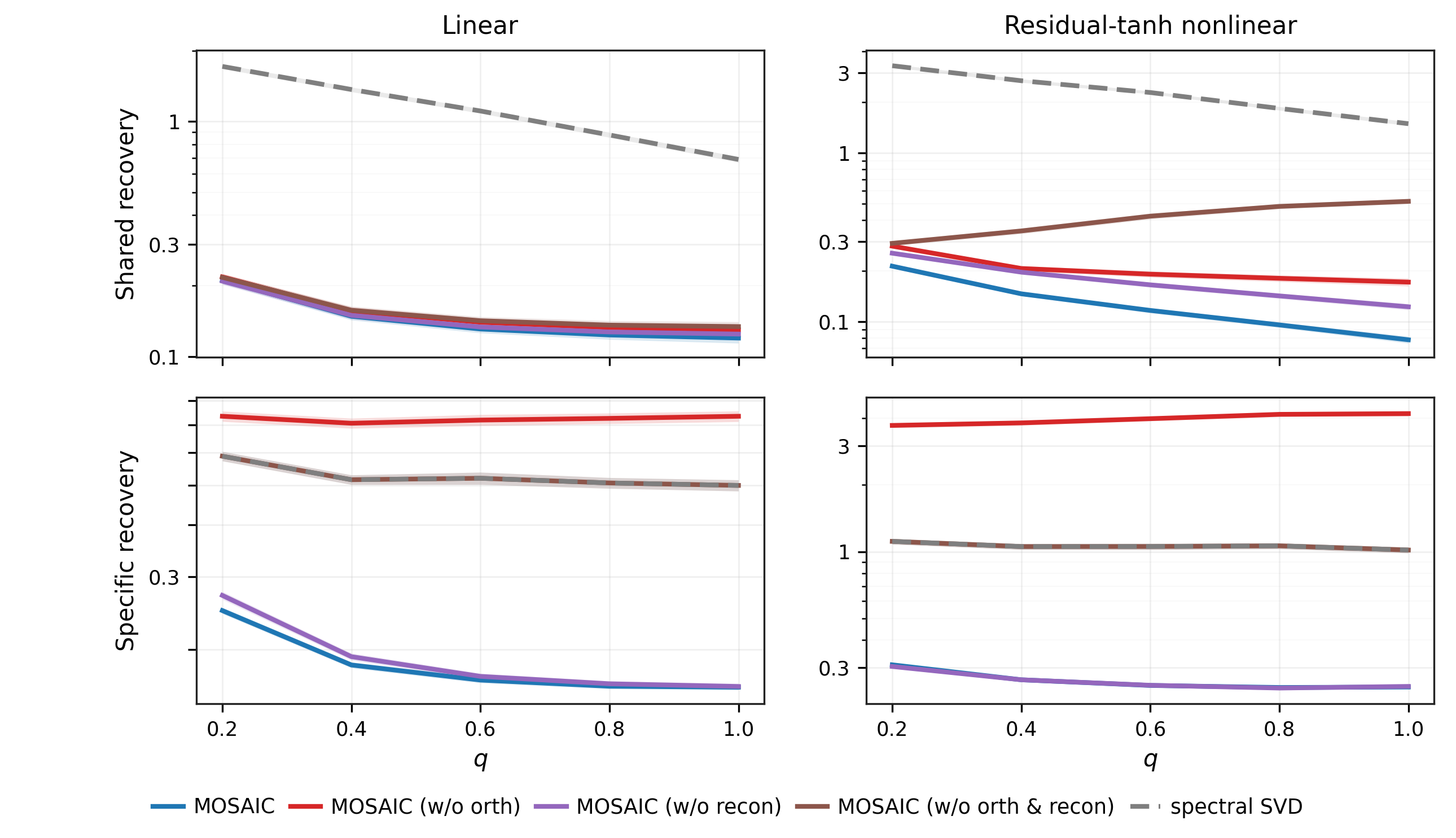}
\caption{Representation recovery as $q$ varies, with $N=5000$. Shaded bands are pointwise 95\% Monte Carlo confidence intervals for the mean over 100 replications.}
\label{fig:exp1-q-sweep}
\end{figure}

Figure~\ref{fig:exp1-sample-size-recovery} shows that representation recovery generally improves as $N$ increases, with the nonlinear observation setting being substantially harder than the linear setting. The main separation among methods appears in modality-specific recovery. Without orthogonal decoupling, the specific representation tends to retain variation that should be attributed to the shared representation; without reconstruction, the learned representation is less effective at preserving modality-specific information after alignment. Overall, MOSAIC is the most stable across the two recovery targets.

Figure~\ref{fig:exp1-q-sweep} shows a complementary pattern as the modality observation probability $q$ varies. Larger $q$ improves recovery because more subjects have co-observed modalities, which strengthens the empirical alignment signal. This benefit is clearest for methods that can use alignment without collapsing the representation structure. The spectral SVD baseline is limited by its linear covariance construction, while the ablations remain more sensitive to leakage between shared and modality-specific components. Optimization and tuning details are provided in Supplementary Section~E.1.

\subsection{Experiment 2: Prediction under Missingness Patterns}
\label{sec:simulation-exp2}

Experiment~2 evaluates downstream prediction after representation learning. We first compare MOSAIC with missing-modality learning baselines under the same shared and modality-specific latent structure as Experiment~1. We then compare MOSAIC with the sharing pattern submodels (SPSM) estimator of \citet{stempfle2023sharing}, applied to the same aligned representations used by MOSAIC; we refer to this baseline as Aligned-SPSM. We also use four internal variants to assess overlap borrowing, focal-pattern calibration, pooled fitting, and representation learning. We maintain $L=3$ modalities, and the covariate generation settings for the linear and nonlinear regimes follow Experiment~1. Responses are generated from the complete latent representation with additive Gaussian noise; detailed coefficient generation and training settings are in Supplementary Section~E.2. 

The external baselines are M3Care \citep{zhang2022m3care}, MedFuse \citep{hayat2022medfuse}, ShaSpec \citep{wang2023shaspec}, IF-MMIN \citep{zhao2021missing,zuo2023exploiting}, TFR-Net \citep{yuan2021transformer}, and CorrKD \citep{corrkd2024}.  Brief descriptions of the comparison methods are provided in Supplementary Section~E.11. All methods use the same missingness indicators $m_i^{(l)}$, train-validation split, and early-stopping rule within each replication. We fix $N=5000$ and vary $n$. For evaluation, we generate a balanced test set with 200 subjects from each nonempty missingness pattern. We first compute the test MSE within each pattern, denoted by $\mathrm{MSE}_k$, and report the equally weighted average $\mathrm{MSE}_{\mathrm{bal}}=K^{-1}\sum_{k=1}^K\mathrm{MSE}_k$ to prevent common patterns from dominating the metric; with $L=3$, this gives $K=7$ patterns.

\begin{figure}[!t]
\centering
\includegraphics[width=\textwidth]{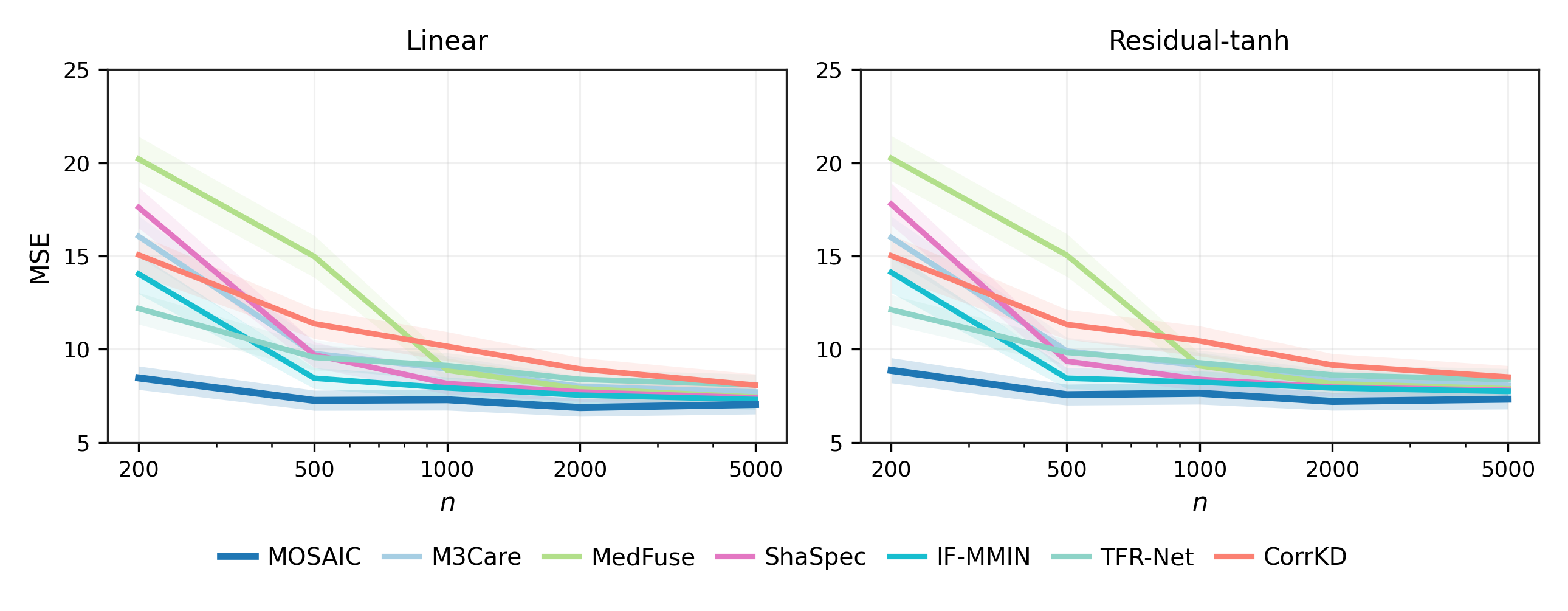}
\caption{Comparison between MOSAIC and external baselines in Experiment~2. Shaded bands are pointwise 95\% Monte Carlo confidence intervals for the mean over 100 replications.}
\label{fig:exp2-ours-vs-external}
\end{figure}

Figure~\ref{fig:exp2-ours-vs-external} shows that MOSAIC achieves the lowest balanced-pattern MSE in both observation settings. IF-MMIN is the strongest external competitor, but the gap remains visible across the labelled sample-size range. The comparison supports the value of pattern-adaptive prediction rather than treating missing modalities only through fusion or reconstruction.

For the comparison, we use a residual-dependence setting with $q=0.6$, where the shared representation remains independent but modality-specific representations become correlated through a common residual factor. We vary $\rho\in\{0,0.3,0.6\}$, which increases the calibration gap. The complete setting is provided in Supplementary Section~E.2. We compare MOSAIC with Aligned-SPSM and four internal variants. By construction, Aligned-SPSM uses the same learned shared and modality-specific representations as MOSAIC, replacing the overlap-and-calibration head with a shared coefficient vector and sparse pattern-specific deviations. Its penalties are selected jointly by balanced-pattern validation MSE. MOSAIC-local uses only focal-pattern samples. MOSAIC-pooled ignores pattern structure by fitting one predictor on zero-filled representations from all patterns. MOSAIC (w/o calib) keeps the overlap-based preliminary predictor but removes the focal-pattern calibration step. MOSAIC (w/o rep) replaces the learned  representations with modality-wise PCA features.

\begin{figure}[!t]
\centering
\includegraphics[width=\textwidth]{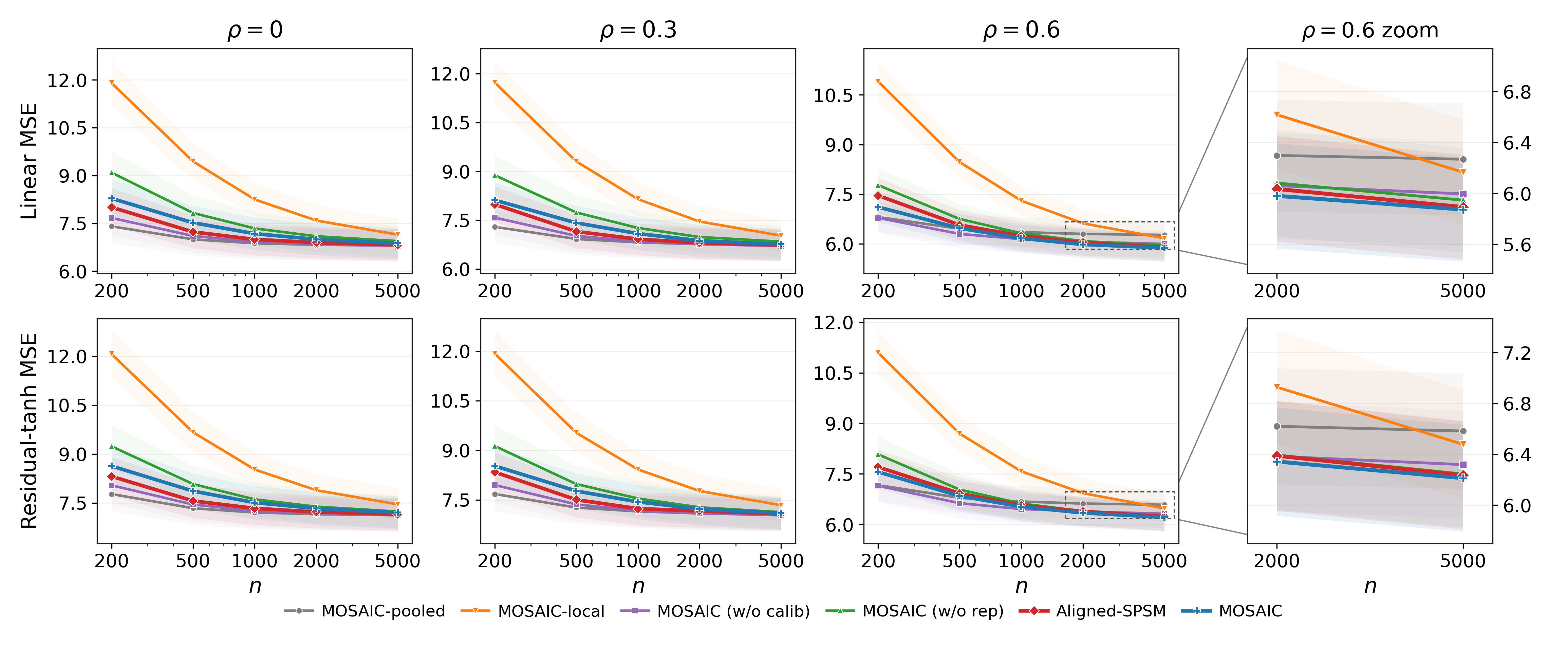}
\caption{Downstream decomposition as residual dependence among modality-specific representations varies. Shaded bands are pointwise 95\% confidence intervals for the mean over 100 replications. The right column enlarges the $\rho=0.6$ results at $n\in\{2000,5000\}$.}
\label{fig:exp2-downstream-rho-sweep}
\end{figure}

The comparison with Aligned-SPSM shows a crossover rather than uniform dominance (Figure~\ref{fig:exp2-downstream-rho-sweep}). Aligned-SPSM has lower mean MSE at every evaluated $n$ when $\rho=0$ and $\rho=0.3$, although the difference becomes small at the largest $n$ under $\rho=0.3$. At $\rho=0.6$, MOSAIC has lower mean MSE at every $n$ in both observation regimes. This reversal is consistent with coefficient sharing being effective when pattern-specific discrepancies are small, whereas overlap-based prediction followed by focal-pattern calibration becomes advantageous as residual dependence and rule mismatch increase.

The internal variants provide complementary evidence. MOSAIC has lower mean MSE than MOSAIC-local and MOSAIC (w/o rep) in all evaluated settings. At $\rho=0.6$, MOSAIC overtakes MOSAIC-pooled from $n=1000$ and MOSAIC (w/o calib) from $n=2000$ in both observation regimes, while these simpler variants remain competitive at lower residual dependence or smaller sample sizes. Thus, the results support the value of overlap borrowing throughout, while the benefit of calibration emerges when both the rule mismatch and the focal-pattern sample size are sufficiently large. Additionally, we compare MOSAIC with the raw-imputation baseline and extend the analysis in Supplementary Section~E.2.

\section{Real Data Applications}\label{sec real data}

We evaluate MOSAIC on next-visit intensive care unit (ICU) mortality prediction using the Medical Information Mart for Intensive Care (MIMIC) database \citep{johnson2023mimic}. MIMIC contains longitudinal structured codes, clinical notes, and medical images, but these modalities are not available for every patient. These naturally occurring modality-availability patterns arise from routine clinical workflows, providing a real-world blockwise missingness setting for evaluation. Additional details on cohort construction, modality-specific preprocessing, training protocol, dataset statistics, and empirical observed-modality-pattern distributions are provided in Supplementary Sections~E.3 and E.5--E.8.

To examine performance under other tasks, we also evaluate MOSAIC under controlled modality masking on IEMOCAP emotion recognition with audio and text \citep{Busso2008} and FairDomain glaucoma classification with two ophthalmic imaging modalities \citep{tian2025fairdomain}. Full results and analyses are reported in Supplementary Section~E.9.

\subsection{MIMIC Next-Visit ICU Mortality Prediction}

We formulate the MIMIC application as a next-visit ICU mortality prediction task. To construct this task, we restrict the cohort to patients with at least two ICU visits, yielding 130{,}873 patients in total. We then split this cohort at the patient level into fixed training, validation, and test sets in a $7:1:2$ ratio. 
The validation set is used for model selection, and the test set is reserved for final evaluation. 
For patient $i$, let $T_i$ denote the number of observed ICU visits; the covariate history consists of visits $1,\ldots,T_i-1$, and the outcome indicates whether mortality occurs at visit $T_i$. The final visit is used only to define the outcome and is excluded from the covariate history.

We construct covariates from three modalities. The structured-code covariates contain diagnosis, medication, diagnosis-related group (DRG), and laboratory event streams from historical visits. The image covariates contain a chest X-ray when available \citep{PhysioNet-mimic-cxr-jpg-2.1.0}. The note covariates contain discharge notes and radiology notes from historical visits. A modality is coded as missing if no historical information from that modality is available.

In addition to overall test performance, we report performance within observed-modality patterns to assess whether the observed gains persist across the prediction tasks induced by different modality-availability patterns. To avoid noisy pattern-level estimates, we display only patterns with at least 20 test samples and at least one positive test case. The retained MIMIC patterns include code-only, code+note, and code+note+image configurations; detailed sample sizes for each observed-modality pattern are reported in Supplementary Section~E.8.

\begin{figure}[ht]
\centering
\includegraphics[width=\textwidth]{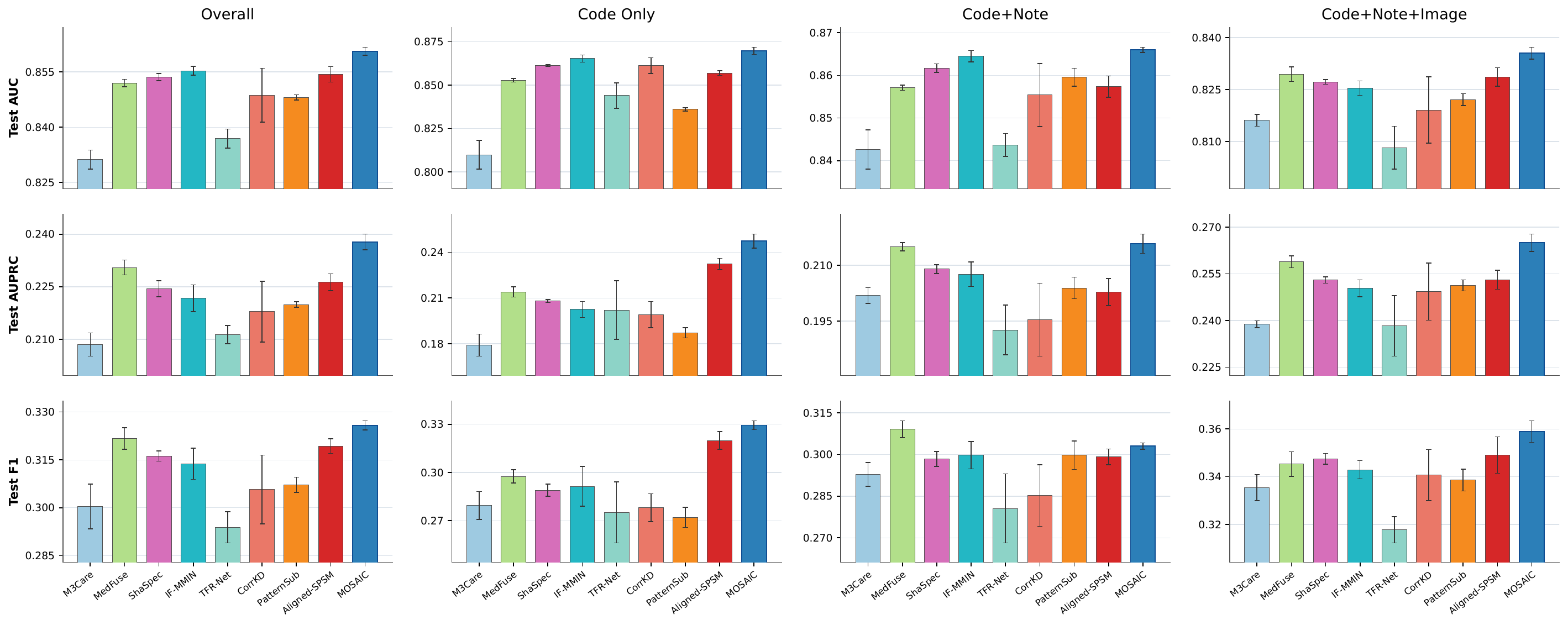}
\caption{Performance on MIMIC next-visit ICU mortality prediction. Bars show mean performance over five seeds, with error bars indicating standard deviations.}
\label{fig:real-mimic-patterns}
\end{figure}

We compare MOSAIC with the comparison methods introduced in Section~\ref{sec:simulation-exp2}, together with the pattern-submodel baseline PatternSub \citep{mercaldo2020missing}. Figure~\ref{fig:real-mimic-patterns} reports overall and pattern-specific test performance over five random seeds.

MOSAIC achieves the highest overall mean area under the receiver operating characteristic curve (AUROC), area under the precision--recall curve (AUPRC), and F1 score among the evaluated methods, with values of $0.861$, $0.238$, and $0.326$, respectively. In the code-only pattern, the separation between MOSAIC and the comparison methods is generally larger than in the other modality-availability patterns. This behaviour is consistent with cross-pattern borrowing followed by focal-pattern correction: observations from richer patterns can contribute to the code-only predictor through the shared representation and the code-specific representation, both of which are available from code histories. In the code+note+image pattern, MOSAIC again has the highest mean AUROC, AUPRC, and F1, showing that its advantage is retained when all three modalities are observed. By contrast, the leading methods are more closely grouped in the code+note pattern, where the advantage of MOSAIC is less pronounced than in the other two patterns. Overall, MOSAIC performs favourably across heterogeneous modality-availability patterns, although the magnitude of its advantage varies with the observed modalities.

\subsection{Component Analysis on the MIMIC Application}

We further compare MOSAIC with the four internal variants from Section~\ref{sec:simulation-exp2} to isolate the contributions of cross-pattern borrowing, representation learning, calibration, and pattern-specific final prediction.

\begin{figure}[ht]
\centering
\includegraphics[width=\textwidth]{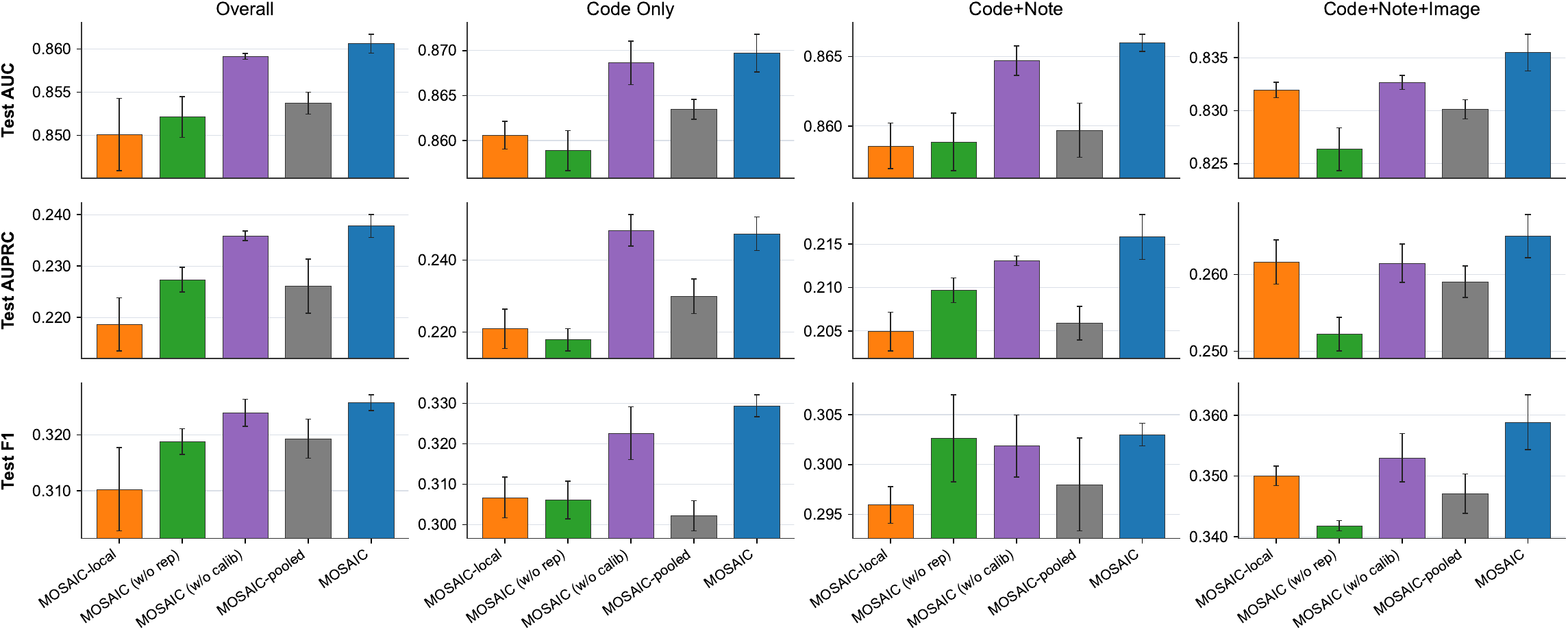}
\caption{Component analysis of MOSAIC on MIMIC next-visit ICU mortality prediction. MOSAIC-local removes cross-pattern borrowing, MOSAIC (w/o rep) removes the representation-learning component, MOSAIC (w/o calib) removes focal-pattern calibration, and MOSAIC-pooled removes pattern-specific final prediction. Bars show mean performance over five random seeds, with error bars indicating standard deviations.}

\label{fig:real-component-analysis}
\end{figure}

Figure~\ref{fig:real-component-analysis} reports the component analysis. The full model achieves the best performance and also leads in most pattern-specific comparisons, although the size of the differences varies across availability patterns. Its gains over MOSAIC (w/o rep) highlight the value of the learned shared and modality-specific representations, while its gains over MOSAIC (w/o calib) indicate that focal-pattern correction provides an additional refinement. The largest differences between the full model and MOSAIC-local or MOSAIC-pooled occur in the code-only group. With neither notes nor images available, a local predictor cannot use observations from patterns with additional modalities. The comparison with MOSAIC-local reflects the value of cross-pattern borrowing, while the comparison with MOSAIC-pooled shows the benefit of pattern-specific final prediction.

The improvement of the full model over MOSAIC-local is smaller in the code+note+image group. This is expected, as this group already contains the richest observed information, leaving less scope for borrowing from other patterns to improve prediction. In the code+note pattern, the differences between MOSAIC and its components are more modest. This is consistent with the pattern-specific results in Figure~\ref{fig:real-mimic-patterns}, where several strong methods have similar performance in this observed-modality configuration. 


\section{Conclusion}\label{sec conclusion}

MOSAIC borrows information across multimodal missingness patterns while retaining a separate prediction rule for each pattern. Once modalities are aligned using co-observed data, observations from patterns with no raw modality in common can still contribute through the shared representation, while focal-pattern calibration accounts for the remaining rule discrepancy. The theory and experiments show that the gain over local fitting depends on representation coverage, the pattern-specific sample size, and the size of the required correction. The current guarantees rely on a linear latent-representation model and MCAR; extending them to nonlinear representations and more general missingness mechanisms remains open.






{\footnotesize

\bibliography{references.bib}

}

\end{document}